\RequirePackage{fix-cm}
\documentclass[smallextended]{svjour3}
\smartqed
\usepackage{graphicx,epsf,epsfig,amssymb}
\usepackage{amsmath,amssymb}
\usepackage{amsfonts}
\usepackage{xspace} 
\usepackage[usenames]{color}
\usepackage{dcolumn}
\usepackage{bm}
\usepackage{mathrsfs}
\usepackage{aas_macros} 
\usepackage[colorlinks=true]{hyperref}
\usepackage[all]{hypcap} 
\usepackage[utf8]{inputenc} 
\usepackage{cite}
\usepackage[normalem]{ulem}
\newcommand{\eq}{\begin{equation}}
\newcommand{\be}{\begin{equation}}
\newcommand{\eeq}{\end{equation}}
\newcommand{\ee}{\end{equation}}
\newcommand\ba{\begin{eqnarray}}
\newcommand\ea{\end{eqnarray}}

\usepackage[table]{xcolor}
\usepackage[usenames]{color}
\usepackage{multirow}

\newcommand{\msun}{\ensuremath{M_\odot}}

\definecolor{darkred}{rgb}{0.8, 0.0, 0.0}
\definecolor{midred}{rgb}{0.94, 0.19, 0.22}
\definecolor{litered}{rgb}{0.91, 0.33, 0.5}
\definecolor{superlitered}{rgb}{0.99, 0.56, 0.67}
\definecolor{darkblue}{rgb}{0.0, 0.0, 0.55}
\definecolor{midblue}{rgb}{0.12, 0.56, 1.0}
\definecolor{liteblue}{rgb}{0.0, 0.72, 0.92}
\definecolor{superliteblue}{rgb}{0.49, 0.98, 1.0}
\definecolor{darkgreen}{rgb}{0.09, 0.45, 0.27}
\definecolor{midgreen}{rgb}{0.52, 0.73, 0.4}

\newcommand{\gcel}{{\cellcolor{midgreen}}}
\newcommand{\rcel}{{\cellcolor{midred}}}
\newcommand{\ycel}{{\cellcolor{yellow}}}
\newcommand{\bcel}{{\cellcolor{midblue}}}
\newcommand{\wcel}{{\cellcolor{white}}}

\newcommand{\Tobs}{T_{\rm data}}
\newcommand{\Telapse}{T_{\rm elapsed}}

\begin{document}

\title{The Effect of Mission Duration on LISA Science Objectives}

\titlerunning{}

\author{Pau Amaro Seoane \and
        Manuel Arca Sedda \and 
        Stanislav Babak \and
        Christopher P.\ L.\ Berry \and 
        Emanuele Berti \and
        Gianfranco Bertone \and
        Diego Blas \and
        Tamara Bogdanovi\'c \and
        Matteo Bonetti \and
        Katelyn Breivik \and
        Richard Brito \and
        Robert Caldwell \and
        Pedro R. Capelo \and
        Chiara Caprini \and
        Vitor Cardoso \and
        Zack Carson \and
        Hsin-Yu Chen \and
        Alvin J. K. Chua \and
        Irina Dvorkin \and
        Zoltan Haiman \and
        Lavinia Heisenberg \and
        Maximiliano Isi \and
        Nikolaos Karnesis \and
        Bradley J. Kavanagh \and
        Tyson B. Littenberg \and
        Alberto Mangiagli \and
        Paolo Marcoccia \and
        Andrea Maselli \and
        Germano Nardini \and
        Paolo Pani \and
        Marco Peloso \and
        Mauro Pieroni \and
        Angelo Ricciardone \and
        Alberto Sesana \and
        Nicola Tamanini \and
        Alexandre Toubiana \and
        Rosa Valiante \and
        Stamatis Vretinaris \and
        David J. Weir \and
        Kent Yagi \and
        Aaron Zimmerman
        }
        
\institute{
           Pau Amaro Seoane \at
           Institute of Multidisciplinary Mathematics, Universitat Polit{\`e}cnica de Val{\`e}ncia, Spain\\
           DESY Zeuthen, Germany\\
           Institute of Applied Mathematics, Academy of Mathematics and Systems Science, CAS, Beijing, China\\
           Kavli Institute for Astronomy and Astrophysics, Beijing, China
           \and
           Manuel Arca Sedda \at
           Astronomisches Rechen-Institut, Zentr\"um f\"ur Astronomie, Universit\"at Heidelberg, M\"onchofstr. 12-14, Heidelberg, Germany
           \and
           Stanislav Babak \at
           Universit\'e de Paris, CNRS, Astroparticule et Cosmologie, F-75006 Paris, France\\
           Moscow Institute of Physics and Technology, Dolgoprudny, Moscow region, Russia
           \and
           Christopher P.\ L.\ Berry \at
           Center for Interdisciplinary Exploration and Research in Astrophysics (CIERA), Department of Physics and Astronomy, Northwestern University, 1800 Sherman Ave, Evanston, IL 60201, USA\\
           SUPA, School of Physics and Astronomy, University of Glasgow, Kelvin Building, University Ave, Glasgow G12 8QQ, UK
           \and
           Emanuele Berti \at
           Department of Physics and Astronomy, Johns Hopkins University, 3400 N. Charles St, Baltimore, Maryland 21218, USA
           \and
           Gianfranco Bertone\at
           Gravitation and Astroparticle Physics in Amsterdam (GRAPPA), and Institute for Theoretical Physics, University of Amsterdam, Science Park 904, 1098 XH Amsterdam, The Netherlands
           \and
           Diego Blas \at
           Theoretical Particle Physics and Cosmology Group, Department of Physics, King's College London, Strand, WC2R 2LS London, UK\\
           Grup  de  F\'isica  Te\`orica,  Departament  de  F\'isica, \\ Universitat  Aut\`onoma  de  Barcelona,  08193  Bellaterra, Spain\\
           Institut de Fisica d’Altes Energies (IFAE), \\The Barcelona Institute of Science and Technology, Campus UAB, 08193 Bellaterra , Spain
           \and
           Tamara Bogdanovi\'c \at
           Center for Relativistic Astrophysics and School of Physics, Georgia Institute of Technology, Atlanta, GA 30332, USA
           \and 
           Matteo Bonetti \at
           Universit\`a degli Studi di Milano-Bicocca, Piazza della Scienza 3, Milano 20126, Italy
           \and
           Katelyn Breivik \at
           Center for Computational Astrophysics, Flatiron Institute, New York, NY 10010, USA
           \and
           Richard Brito \at
           CENTRA, Departamento de F\'{\i}sica, Instituto Superior T\'ecnico -- IST, Universidade de Lisboa -- UL, Avenida Rovisco Pais 1, 1049 Lisboa, Portugal
           \and
           Robert Caldwell \at
           Department of Physics and Astronomy, HB6127 Wilder Lab, Dartmouth College, Hanover, New Hampshire 03755, USA
           \and
           Pedro R. Capelo \at
           Center for Theoretical Astrophysics and Cosmology, Institute for Computational Science, University of Zurich, Winterthurerstrasse 190, CH-8057 Z\"urich, Switzerland
           \and
           Chara Caprini \at
           Laboratoire Astroparticule et Cosmologie, CNRS UMR 7164, Universit\'e Paris-Diderot, 10 rue Alice Domon et L\'eonie Duquet, 75013 Paris, France
           \and
           Vitor Cardoso \at
           CENTRA, Departamento de F\'{\i}sica, Instituto Superior T\'ecnico -- IST, Universidade de Lisboa -- UL, Avenida Rovisco Pais 1, 1049 Lisboa, Portugal
           \and
           Zack Carson \at
           Department of Physics, University of Virginia, P.O.~Box 400714, Charlottesville, VA 22904-4714, USA
           \and
           Hsin-Yu Chen \at
           LIGO Laboratory, Massachusetts Institute of Technology, Cambridge, Massachusetts 02139, USA
            \and
           Alvin J. K. Chua \at
           Theoretical Astrophysics Group, California Institute of Technology, Pasadena, CA 91125, U.S.A.
           \and
           Irina Dvorkin \at
           Institut d'Astrophysique de Paris, Sorbonne Universit\'{e} \& CNRS, UMR 7095, 98 bis bd Arago, 75014 Paris, France
           \and
           Zoltan Haiman \at
           Department of Astronomy, Columbia University, 550 W. 120th St., New York, NY, 10027, USA
           \and
           Lavinia Heisenberg \at
           Institute for Theoretical Physics, ETH Zurich, Wolfgang-Pauli-Strasse 27, 8093, Z\"urich, Switzerland
           \and
           Maximiliano Isi \at
           LIGO Laboratory, Massachusetts Institute of Technology, Cambridge, Massachusetts 02139, USA
           \and
           Nikolaos Karnesis \at
           Department of Physics, Aristotle University of Thessaloniki, Thessaloniki 54124, Greece\\
           APC, AstroParticule et Cosmologie, Universit\'e de Paris, CNRS, Astroparticule et Cosmologie, F-75013 Paris, France
           \and
           Bradley J. Kavanagh\at Instituto de Física de Cantabria (IFCA, UC-CSIC), Av. de Los Castros s/n, 39005 Santander, Spain
           \and
           Tyson B. Littenberg\at NASA Marshall Space Flight Center, Huntsville, Alabama 35811, USA
           \and
           Alberto Mangiagli \at
           Laboratoire Astroparticule et Cosmologie, CNRS UMR 7164, Universit\'e Paris-Diderot, 10 rue Alice Domon et L\'eonie Duquet, 75013 Paris, France\\
           Department of Physics, University of Milano - Bicocca, Piazza della Scienza 3,I20126 Milano, Italy\\
           National Institute of Nuclear Physics INFN, Milano - Bicocca, Piazza della Scienza 3, 20126 Milano, Italy
           \and
           Paolo Marcoccia \at
           University of Stavanger, N-4036 Stavanger, Norway
           \and
           Andrea Maselli \at
           Gran  Sasso  Science  Institute  (GSSI),  I-67100  L’Aquila,  Italy\\
           INFN, Laboratori Nazionali del Gran Sasso, I-67100 Assergi, Italy
           \and
           Germano Nardini \at
           University of Stavanger, N-4036 Stavanger, Norway
           \and
           Paolo Pani \at
           Dipartimento di Fisica, ``Sapienza" Università di Roma \& Sezione INFN Roma1, Piazzale Aldo Moro 5, 
           00185, Roma, Italy
           \and
           Marco Peloso \at
           Dipartimento di Fisica and Astronomia, Università di Padova \& Sezione INFN Padova, Via Marzolo 8, I-35131 Padova, Italy
           \and
           Mauro Pieroni \at
           Blackett Laboratory, Imperial College London, SW7 2AZ, UK
           \and
           Angelo Ricciardone \at
           1Dipartimento di Fisica e Astronomia ``G. Galilei'', Universit\'a degli Studi di Padova, via Marzolo 8, I-35131 Padova, Italy
           \and
           Alberto Sesana \at
           Department of Physics, University of Milano - Bicocca, Piazza della Scienza 3,I20126 Milano, Italy
           \and
           Nicola Tamanini \at
           Laboratoire des 2 Infinis - Toulouse (L2IT-IN2P3), Universit\'e de Toulouse, CNRS, UPS, F-31062 Toulouse Cedex 9, France
           \and
           Alexandre Toubiana \at
           Universit\'e de Paris, CNRS, Astroparticule et Cosmologie, F-75006 Paris, France\\
           Institut d’Astrophysique de Paris, CNRS \& Sorbonne Universit\'es, UMR 7095, 98 bis bd Arago, 75014 Paris, France
           \and
           Rosa Valiante \at
           INAF-Osservatorio Astronomico di Roma, via di Frascati 33, I-00078 Monteporzio Catone, Italy\\
           INFN, Sezione di Roma I, P.le Aldo Moro 2, I-00185 Roma, Italy
           \and
           Stamatis Vretinaris \at
           Department of Physics, Aristotle University of Thessaloniki, University Campus, 54124, Thessaloniki, Greece
           \and
           David J. Weir \at
           Department of Physics and Helsinki Institute of Physics, PL 64, FI-00014 University of Helsinki, Finland\\
           School of Physics and Astronomy, University of Nottingham, Nottingham NG7 2RD, U.K.
           \and
           Kent Yagi \at
           Department of Physics, University of Virginia, P.O.~Box 400714, Charlottesville, VA 22904-4714, USA
           \and
           Aaron Zimmerman \at
           Center for Gravitational Physics, University of Texas at Austin, Austin, TX 78712, USA
}

\date{Received: date / Accepted: date}

\maketitle

\begin{abstract}
The science objectives of the LISA mission have been defined under the implicit assumption of a 4-yr continuous data stream. Based on the performance of LISA Pathfinder, it is now expected that LISA will have a duty cycle of $\approx 0.75$, which would reduce the effective span of usable data to 3~yr. This paper reports the results of a study by the LISA Science Group, which was charged with assessing the additional science return of increasing the mission lifetime. We explore various observational scenarios to assess the impact of mission duration on the main science objectives of the mission. We find that the science investigations most affected by mission duration concern the search for seed black holes at cosmic dawn, as well as the study of stellar-origin black holes and of their formation channels via multi-band and multi-messenger observations. We conclude that an extension to 6~yr of mission operations is recommended. 

\end{abstract}

\clearpage

\section{Introduction}
\label{sec:intro}

The Laser Interferometer Space Antenna (LISA) \cite{2017arXiv170200786A}\footnote{All acronyms are defined in Table~\ref{tab:acronyms}.} is a space-borne gravitational wave (GW) observatory selected to be ESA's third-large class mission, addressing the science theme of the Gravitational Universe \cite{2013arXiv1305.5720E}. It consists of three spacecraft trailing the Earth around the Sun in a triangular configuration, with a mutual separation between spacecraft pairs of about 2.5 million kilometres. The laser beams connecting the three satellites are combined via time delay interferometry (TDI) \cite{1999ApJ...527..814A} to construct an equivalent pair of two Michelson interferometers.
Thanks to its long armlength, LISA will be most sensitive in the millihertz frequency regime, which is anticipated to be the richest in terms of astrophysical (and possibly cosmological) GW sources, including coalescing massive black hole binaries (MBHBs) across the Universe, millions of binaries of compact objects within our Milky Way, and stochastic GW backgrounds (SGWBs) produced in the early Universe (see Ref.~\cite{2013arXiv1305.5720E,2013GWN.....6....4A} and references therein).

The science objectives (SOs) and science investigations (SIs) of the LISA mission have been defined under the implicit assumption of a 4-yr continuous stream of data, implying that during mission operations, the downtime of the detector is negligible compared to the effective time of data taking. If we define $T_{\rm elapsed}$ to be the time of mission operation (from first light to final shut down) and $T_{\rm data}$ to be the total time of effective data taking, then one can define a duty cycle ${\cal D}=T_{\rm data}/T_{\rm elapsed}\leq 1$. The LISA proposal assumed a duty cycle ${\cal D}>0.95$ \cite{2017arXiv170200786A}. Based on the performance of LISA Pathfinder (which started scientific operations on March 8, 2016 and took data for almost sixteen months), it is now expected that LISA will have a duty cycle ${\cal D}\approx 0.75$, which, for a 4-yr mission, reduces the effective span of usable data to 3~yr.

As we move towards mission adoption by ESA, it is necessary to define a mission design that will fulfill the SOs spelled out in the LISA Science Requirements Document (SciRD)~\cite{SciRD}. In particular, it is of paramount importance to consider the actual condition of data taking and processing, including a realistic duty cycle.
In this study we answer the following questions: are the SOs formulated assuming a 4-yr continuous data stream still achieved with a duty cycle ${\cal D}=0.75$? If they are not, can we achieve them through an extension of the mission duration with the same duty cycle ${\cal D}=0.75$?

Under the assumption of a duty cycle significantly smaller than ${\cal D}=1$, some confusion can arise in the definition of mission duration. Therefore, we start by clarifying the conventions adopted in this study:
\begin{itemize}
    \item $T_{\rm elapsed}$ denotes the nominal mission duration, i.e. the time elapsed since LISA is first turned on, until it is turned off for the last time. The LISA SciRD~\cite{SciRD} assumed $T_{\rm elapsed}=4$~yr.
    \item $T_{\rm data}$ denotes the actual length of the usable data stream. If we have a duty cycle ${\cal D}$, then $T_{\rm data}={\cal D}\times{T_{\rm elapsed}}$. The current best estimate is $T_{\rm data}=3$~yr, given the estimated ${\cal D}=0.75$.
    \item $T_{\rm signal}$ is the typical lifetime of a specific signal in band. Depending on whether this is longer or shorter than $T_{\rm elapsed}$, sources are affected by mission duration in different ways.
\end{itemize}
According to the above definitions, the LISA proposal SciRD assumed ${\cal D}=1$, corresponding to $T_{\rm elapsed}=T_{\rm data}=4$~yr.

In this paper we investigate the potential science impact of increasing the current lifetime of the LISA mission by considering the following scenarios:
\begin{itemize}

\item {\bf SciRD}:
The SciRD configuration from the LISA proposal, i.e. $T_{\rm elapsed}=4$~yr with ${\cal D}=1$. 

\item {\bf T4C}:
Continuous data for 3~yr ($T_{\rm elapsed}=4$~yr with ${\cal D}=0.75$, the current baseline);

\item {\bf T5C}:
Continuous data for 3.75~yr ($T_{\rm elapsed}=5$~yr with ${\cal D}=0.75$);

\item {\bf T6C}:
Continuous data for 4.5~yr ($T_{\rm elapsed}=6$~yr with ${\cal D}=0.75$).
\end{itemize}
The above scenarios can be thought as if there were only a single long gap in the data lasting $(1-{\cal D})\times T_{\rm elapsed}$, occurring either before or after a continuous stretch of data taking.

Besides these continuous-data scenarios, we will also consider scenarios where the $(1-{\cal D})\times T_{\rm elapsed}$ downtime is distributed in short-duration gaps. Assuming that
the gaps have a probability distribution $p(T) = r \exp(-r T)$, such that the expected time between gaps is $\langle T \rangle = \int \mathrm{d}T\, {Tp(T)}=1/r$, we can define several gapped scenarios depending on the rate $r$ as:\footnote{In all these scenarios the duty cycle must remain ${\cal D}=0.75$. This means that the average spacing between gaps must be three times longer than the chosen gap duration $T_{\rm gap}$. For example if $T_{\rm gap}=1$~day, $1/r=3$~days, and so on.}
\begin{itemize}
\item {\bf T4G5}:
Data for 4~yr with gaps of length 5 days such that 25\% of the data is lost (i.e.\ total data stream duration 3~yr), with the time between gaps $T$ following a distribution with $r=1/(15~\mathrm{days})$;
  
\item {\bf T6G5}:
Data for 6~yr with gaps of length 5 days such that 25\% of the data is lost (i.e.\ total data stream duration 4.5~yr), with the time between gaps distributed with $r=1/(15~\mathrm{days})$;

\item {\bf T4G1}:
Data for 4~yr with gaps of length 1~day such that 25\% of the data is lost (i.e.\ total data stream duration 3~yr), with the time between gaps distributed with $r=1/(3~\mathrm{days})$;

\item {\bf T6G1}:
Data for 6~yr with gaps of length 1 day such that 25\% of the data is lost (i.e.\ total data stream duration 4.5~yr), with the time between gaps distributed with $r=1/(3~\mathrm{days})$.

\end{itemize}
Since the main scope of the study is to assess how a duty cycle ${\cal D}=0.75$ due to the presence of random gaps affects LISA's capabilities to reach its SOs, we have primarily focused on the comparison between Cases T4G5, T4G1, T6G5, and T6G1 and the LISA-proposal assumption of 4~yr of continuous data (SciRD).

The paper is organized as follows. The SOs identified in the SciRD document are divided into three main science investigation domains: astrophysics, cosmology, and fundamental physics. Within astrophysics, we further separate SOs according to the relevant GW sources, and we investigate separately MBHBs (Sec.~\ref{sec:WP1}); stellar-mass compact objects, both in the Milky Way and at cosmological distances (Sec.~\ref{sec:WP2}); and extreme mass-ratio inspirals (EMRIs; Sec.~\ref{sec:WP3}). For cosmology, we consider separately the SOs defining LISA's potential to perform standard sirens-based cosmography (Sec.~\ref{sec:WP4}) and those related to the detection of putative SGWBs of cosmological origin (Sec.~\ref{sec:WP5}). In fundamental physics, we investigate separately LISA's capabilities to constrain dark matter (Sec.~\ref{sec:WP6}), test general relativity (Sec.~\ref{sec:WP7}), and explore the nature of black holes (Sec.~\ref{sec:WP8}). We summarize our main findings in Sec.~\ref{sec:summary}. A detailed mapping of SOs and SIs to the sections of this paper can be found in the summary Table~\ref{tab:I} in Sec.~\ref{sec:summary}.

We caution that our simulations are not always homogeneous across SOs. For some signals (e.g. strictly monochromatic or stochastic), to first order, the important quantity to be considered is $T_{\rm data}$, regardless of the duty cycle. Therefore, in the absence of tools for analyzing data with gaps, we sometimes consider continuous streams of length $T_{\rm data}$. These details are specified case-by-case in each section below. Moreover, when gaps are included in the calculations, those are assumed to be lost chunks of the data stream that only affect the source signal-to-noise ratio (SNR) calculations. In reality, gaps will also modify the properties of the noise, which can in turn further affect detection statistics and parameter reconstruction of specific sources. More detailed parameter estimation studies (adopting e.g., the data analysis techniques developed in Ref.~\cite{Baghi:2019eqo,Baghi:2020ygw}) are beyond the scope of this paper.

\section{Formation, evolution, and electromagnetic counterparts of massive black hole mergers}
\label{sec:WP1}

In this section we consider the impact of the mission lifetime on SOs related to the formation, evolution, and electromagnetic (EM) counterparts of MBHBs. We first examine the effect of the mission lifetime ($T_{\rm elapsed}$) and then focus on the impact of gaps of different length given a duty cycle ${\cal D}=0.75$. Our results will be formulated in terms of three timescales: $T_{\rm signal}$, $T_{\rm elapsed}$, and $T_{\rm data}$.

Most MBHBs stay in the LISA band for a period of time (weeks, at most months) much shorter than LISA's lifetime, hence $T_{\rm signal} \ll T_{\rm elapsed}$. This means that the number of observed sources scales linearly with $T_{\rm elapsed}$. It is therefore important to investigate the effect of gaps of different lengths on the resulting number of detections and compare it to a scenario with a continuous data stream.  We thus focus on comparing the SciRD, T4G1, and T4G5 scenarios, with the understanding that results scale linearly for longer mission duration.

We run the light seed (hereafter popIII, since the seeds originate from Population III stars) and heavy seed models used in Ref.~\cite{2019MNRAS.486.4044B}.\footnote{The heavy seed model is equivalent to the model Q3d in Sec.~\ref{sec:WP4}.} 
The two models describe the co-evolution of MBHBs with their host galaxies %
assuming that MBH progenitors are either light ($\sim 100 \rm$~M$_\odot$; popIII remnants) or heavy ($10^5 \rm M_\odot$) seeds forming at redshifts $15<z<20$. In both models, MBHBs are driven to coalescence via interactions with stars, gas, and/or a third black hole, and the evolution of their orbital eccentricity is followed self-consistently (see Ref.~\cite{2019MNRAS.486.4044B} for details).

Using these fiducial models, in which binary merger timescales (of the order of millions to billions of years) depend on the host galaxy properties,
we first assess the impact of gaps on the overall number of detections. We thus generate a Monte Carlo sample of 100~yr of MBHB mergers and consider either continuous observations or data with 1 day or 5-day gaps resulting in ${\cal D}=0.75$. To assess the global impact of gaps, we divide this set in 25 chunks of 4~yr each and compute the number and SNR distribution of detected systems for the cases SciRD, T4G1, and T4G5. We assume SNR $=8$ as a detection threshold.

\begin{figure*}
\centering
\includegraphics[width=0.47\textwidth]{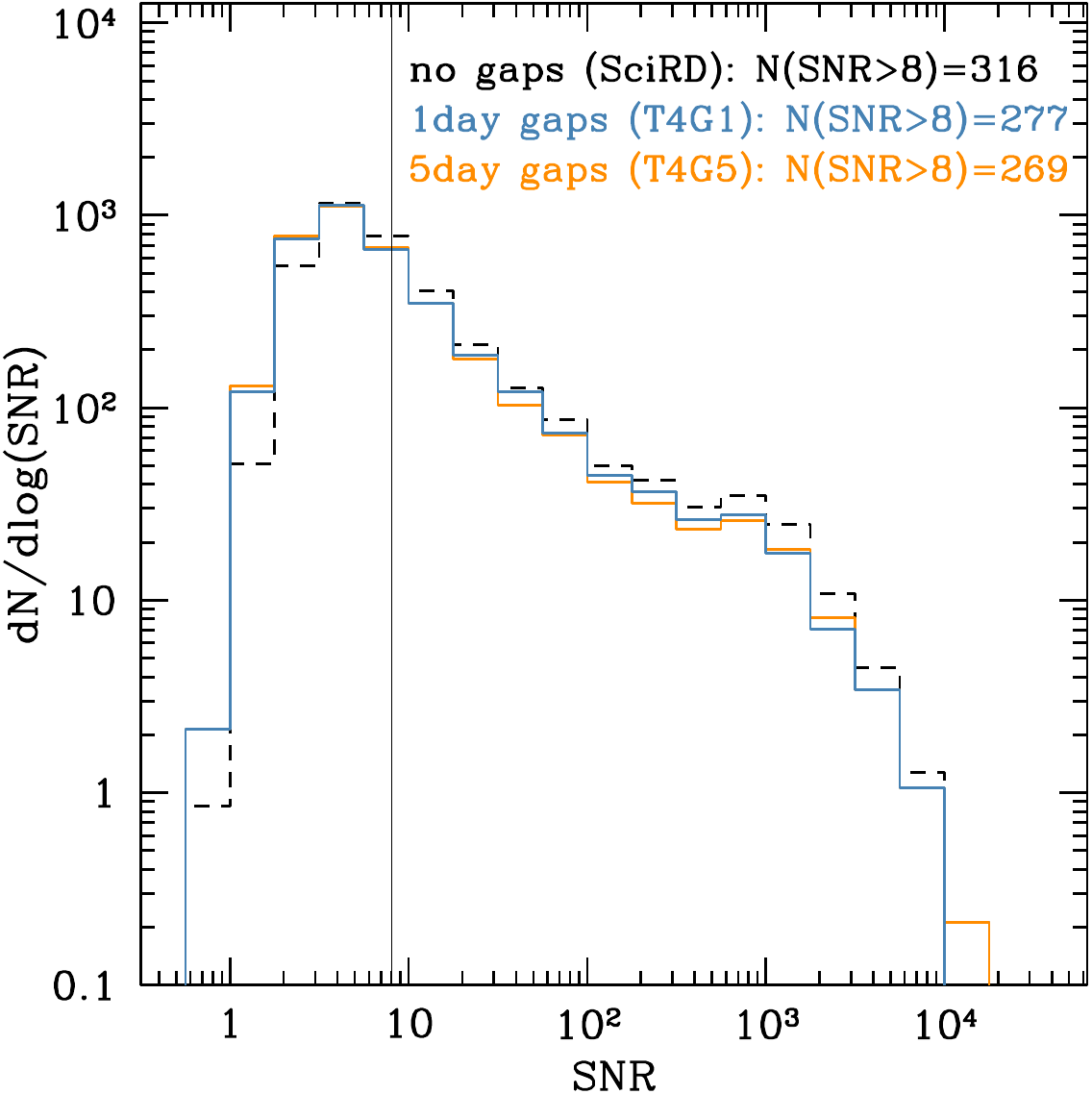}
\includegraphics[width=0.47\textwidth]{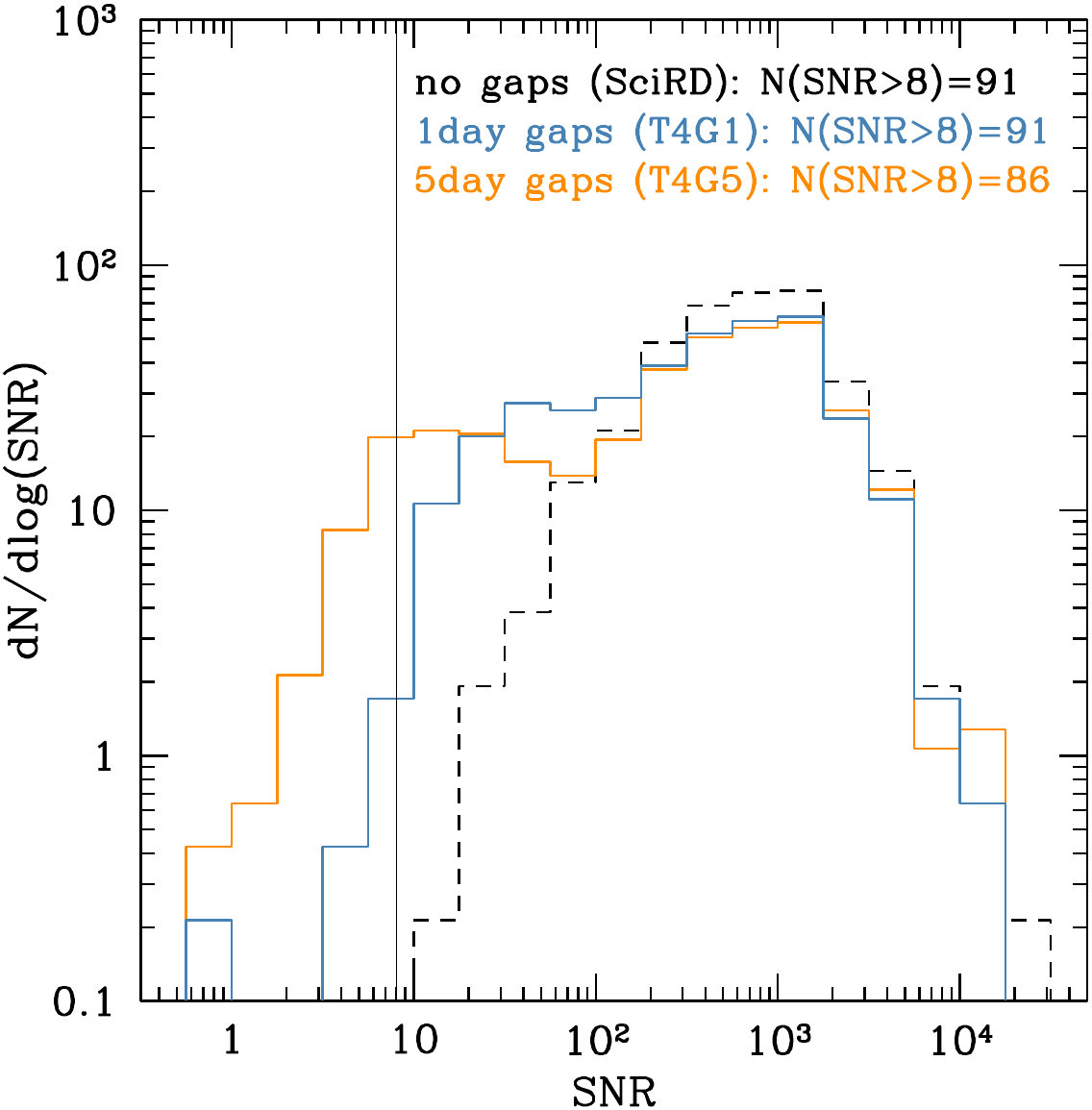}
\caption{SNR distribution for MBHBs in the popIII (left) and heavy seed (right) cases. In both scenarios we assume $T_{\rm elapsed}=4$~yr. We compare observational scenarios with continuous data (SciRD) and data with gaps of different length and ${\cal D}=0.75$ (T4G1, T4G5), as labeled in the figure. The legends show the number of sources observed with SNR~$>8$. 
}
\label{fig:mbhbnumber}
\end{figure*}

\begin{figure}[t]
    \centering
    \includegraphics[width=1.0\textwidth]{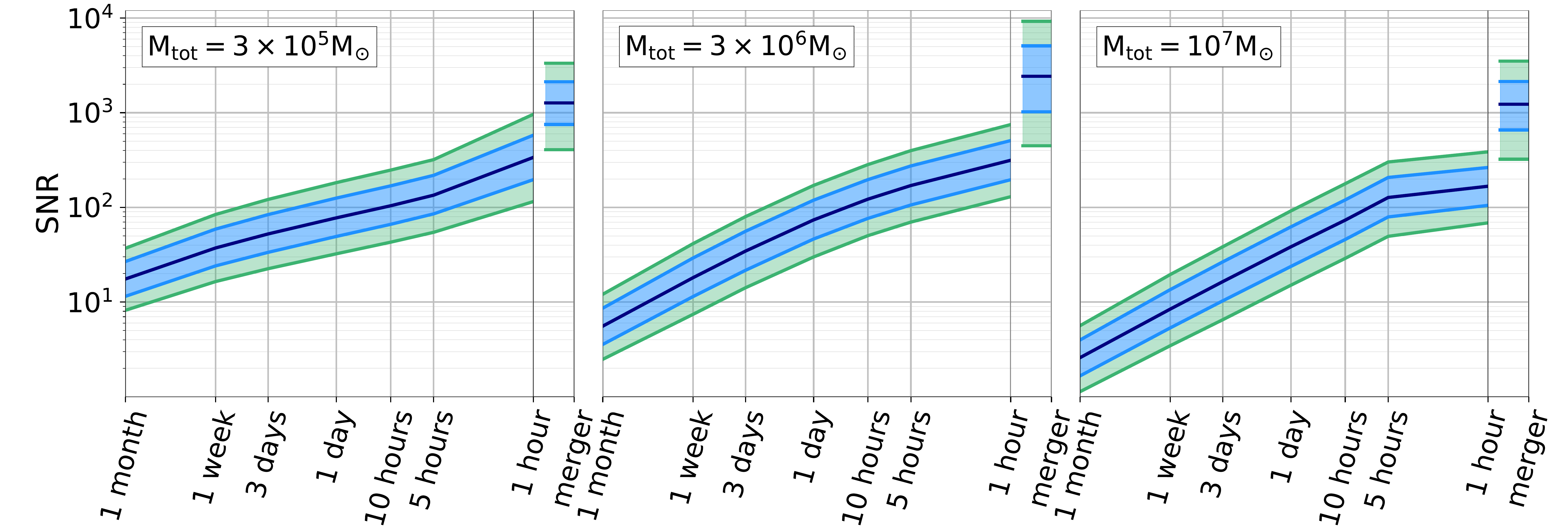}
    \caption{SNR evolution as a function of time to merger for selected MBHBs at $z=1$. In each panel we show the median and the 68\% and 95\% confidence regions for a sample of $10^4$ simulated binaries with the indicated total mass and otherwise randomized parameters (sky location, inclination, polarization, etc.). The mass ratios are randomly drawn in the range $[0.1, 1]$. (Adapted from Ref.~\cite{2020PhRvD.102h4056M}.)}
    \label{fig:SNR}
\end{figure}

The results reported in Fig.~\ref{fig:mbhbnumber} show that the impact of gaps depends on the nature of MBH seeds. In the heavy seeds case, compared to the SciRD scenario, there is no loss of detections ($>99$\% detections) in the T4G1 scenario, whereas in the T4G5 scenario 95\% of the systems are still detected. Gaps have a stronger impact in the popIII case where, compared to SciRD, 88\% and 85\% of the sources are still detected in the T4G1 and T4G5 scenarios, respectively. The first thing to notice is that those fractions are always larger than the 75\% duty cycle. This is because MBHBs stay in band for weeks or more, as shown by the SNR accumulation depicted in Fig.~\ref{fig:SNR} (from Ref.~\cite{2020PhRvD.102h4056M}) for systems of total mass $3 \times 10^5\msun$, $3 \times 10^6\msun$, and $10^7\msun$ at $z = 1$. Random gaps of few days will remove portions of the signal, but in the vast majority of the cases there will still be enough SNR build-up to guarantee detection. This is especially true if gaps are short and sources have high SNR, which is the case for heavy seeds and T4G1. The longer are the gaps and the lower is the typical source SNR, the higher are the chances that sources end up below the detection threshold. This is why gaps are more detrimental if they last 5 days and in the popIII scenario.

Despite introducing a duty cycle has a sub-linear effect on the overall number of detections, there are specific types of sources that might be more severely affected, jeopardizing some of the LISA mission goals. In the following, we focus on the opposite ends of the MBHB spectrum, namely low-mass seeds at high redshift and low-redshift massive systems. Again, we fix $T_{\rm elapsed}=4$~yr and compare configurations SciRD, T4G5, and T4G1.

The number of observed high-redshift ($z>10$), low-mass ($M<10^3\msun$) systems is severely impacted by the presence of gaps reducing the duty cycle to ${\cal D}=0.75$. This is due to a combination of features that are unique to those systems: they are often close to the SNR observability threshold (SNR~$=8$, for MBHBs), they have $T_{\rm signal}\ll T_{\rm elapsed}$, but at the same time $T_{\rm signal}\gg T_{\rm gap}$. Therefore, gaps affect pretty much all of these sources and including gaps in the data causes many of them to drop below the SNR threshold. 

\begin{figure*}\label{fig:MBHB}
    \centering
    \includegraphics[width=0.48\textwidth]{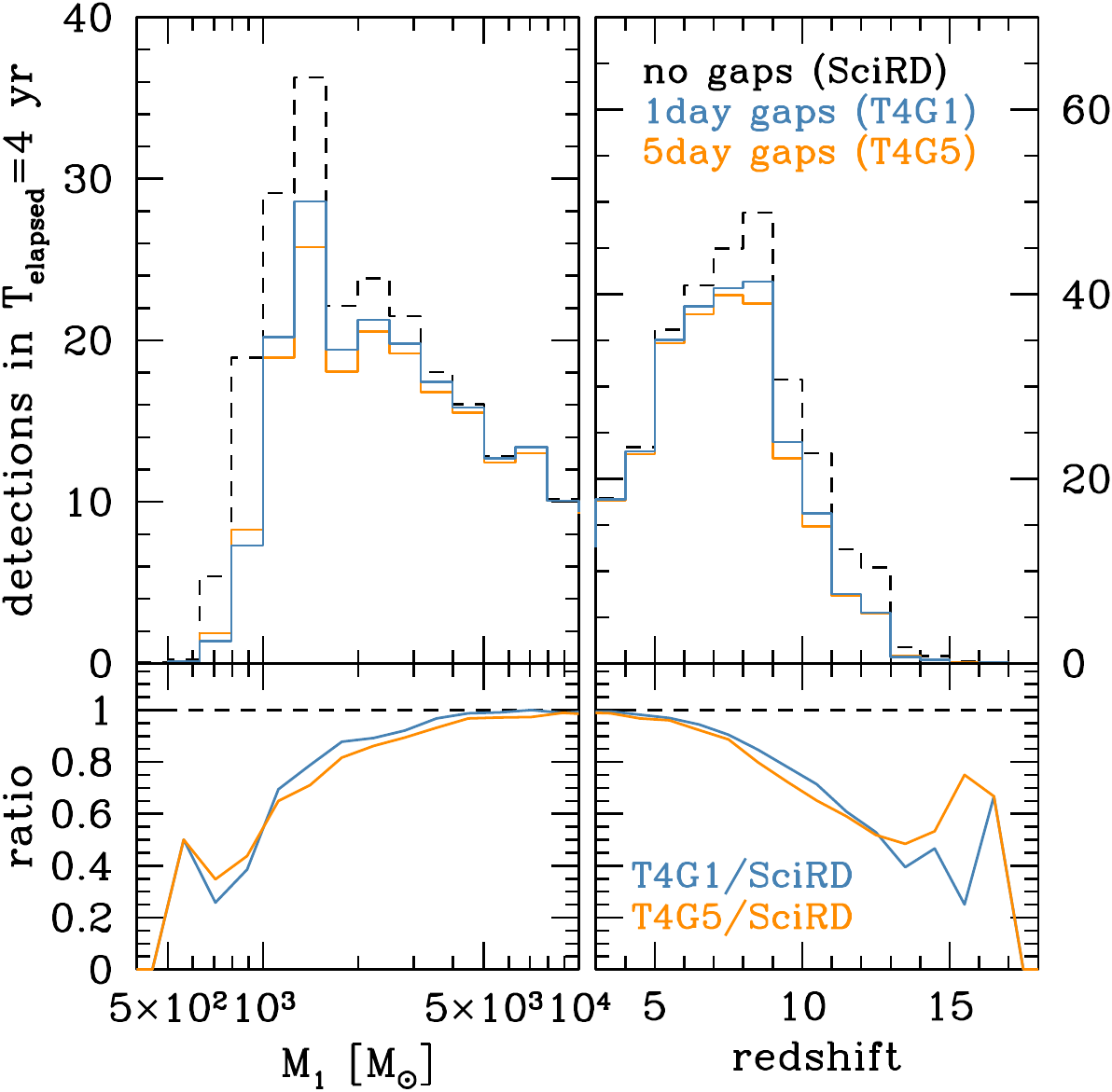}
    \includegraphics[width=0.45\textwidth]{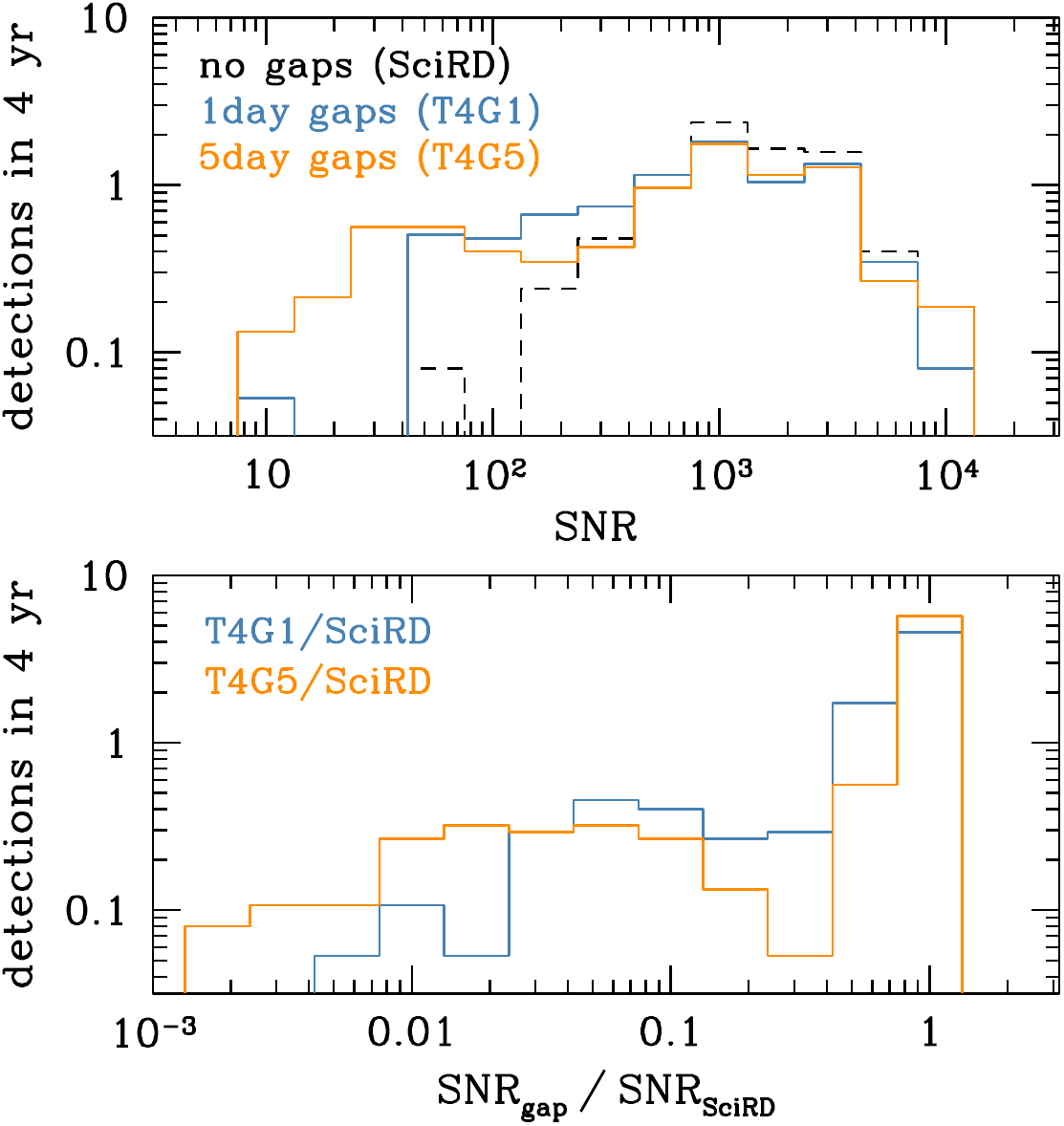}
    \caption{Left: Mass and redshift distribution of sources detected assuming $T_{\rm elapsed}=4$ yr and continuous data ($T_{\rm elapsed}=T_{\rm data}=4$~yr, i.e.\ the SciRD configuration) or data with gaps (T4G5 and T4G1 configurations), as indicated in the figure. The bottom insets show the relative drop in the number of detections for data with gaps compared to the SciRD configuration. Right: The top panel shows the SNR distribution of sources with $M>10^5\msun$ and $z<2$ under the same assumptions made in the left-hand plot. The bottom panel shows the distribution of SNR ratios of continuous observations vs.\ observations with gaps, highlighting the presence of a long tail of sources for which gaps imply a significant loss in SNR.}
    \label{fig:mbhbseed}
\end{figure*}

More specifically, in the SciRD case we expect $\approx 25$ observable sources with $M<10^3\msun$ in the popIII scenario. 
This number drops to $\lesssim 10$ when we consider configurations T4G5 and T4G1, as shown in the left panel of Fig.~\ref{fig:mbhbseed}. These results are qualitatively consistent with the findings of Ref.~\cite{Dey:2021dem}, specifically their Light Seed noSN models, which are similar to the one used here, and the unscheduled gaps scenario with 3-day gaps. For this configuration, Ref.~\cite{Dey:2021dem} finds that the number of observed sources is reduced by $\sim 50\%$ relative to the case without gaps. However, Ref.~\cite{Dey:2021dem} used a more pessimistic gap scenario than the one considered here, which led to an effective duty cycle of  ${\cal D}\simeq 0.65$, compared with ${\cal D}\simeq 0.75$ in our case.

To quantify uncertainties due to model assumptions, we carry out a similar investigation for alternative (more pessimistic) popIII seed models including supernova feedback and other effects that dramatically reduce the number of potential LISA sources (see Ref.~\cite{2020arXiv200603065B} for details). We find that the number of detected low-mass ($M<10^3\msun$) systems drops from $\approx 10$ in the SciRD case to $\lesssim 6$ in the T4G5 and T4G1 scenarios. It is therefore clear that including a 75\% duty cycle into a four year mission operation baseline is severely detrimental to the observation of seed black holes.

At the other end of the MBHB spectrum, several relatively massive ($M>10^5\msun$), nearby ($z<2$) sources might experience a significant SNR drop due to gaps, as shown in the top- and bottom-right panels of Fig.~\ref{fig:mbhbseed}. About 30\% of these sources experience SNR drops by more than a factor of 10. This is more severe for 5-day gaps, in which the merger--ringdown phase of loud signals can be lost entirely. This is emphasized in Fig.~\ref{fig:SNR};
especially for massive systems, the SNR is accumulated in a relatively short period at the end of the binary's lifetime, which can be down to few days only. If the detection threshold is SNR = 8, then 1-day gaps should not affect the detection of any of these systems, whereas 5-day gaps would hinder the detection of some of the more massive binaries with mass above $\sim 10^7\msun$. The sources in the figure are at $z=1$, and increasing the source redshift will inevitably shorten the effective SNR accumulation timescale, exacerbating this potential issue. In practice, this also means that, effectively, a 6-yr mission with 1-day gaps (T6G1) is almost equivalent to a 6-yr mission with $100\%$ duty cycle and no gaps (i.e.\ $T_{\rm elapsed}=T_{\rm data} = 6$~yr), except for a reduced SNR. However, a drop in SNR also carries a penalty, as it implies a proportional deterioration in parameter estimation and (most importantly) sky localization, which might have consequences when searching for EM counterparts.
 
We also expect that gaps will lead to selection effects in terms of certain spin configurations. We did not quantify this bias, but we can make some qualitative considerations. The spin--orbit coupling in spinning black hole binaries can delay (hasten) the onset of the plunge phase compared when the spins are aligned (antialigned) with the orbital angular momentum, respectively. This is often called the orbital hang-up effect~\cite{Campanelli:2006uy}, and it is more pronounced for highly spinning binaries. Therefore, gaps will introduce an observational selection effect: highly spinning binaries with aligned spins will be more likely to be detected relative to other configurations with shorter lifetimes (antialigned, non-spinning, etc.). The highly spinning binaries with aligned spins are also more luminous in GWs, so the two effects would presumably be compounded. This selection effect is expected to be more severe for longer gaps.

Finally, besides considering randomly distributed gaps, which are scheduled or happen without external input, we also propose the following scenario for consideration. Assume a long-lived GW event has already been discovered a month prior to a MBHB merger.  Unfortunately, the SNR is too low, and the source can not be localized on the sky, but at some point well in advance of the merger (e.g., weeks earlier) the merger time can be predicted with an accuracy of a day or so.   Within this final day it can become possible to localize the source, issue alerts, and enable precursor EM observations, or observations of the merger itself. This detection can be  unaffected by gaps if LISA has the capability to adaptively reschedule gaps, when they coincide with the final day of a merger that can be predicted sufficiently in advance. This could significantly mitigate, or eliminate, the deleterious impact of gaps on precursor observations.

These findings have important implications for SO2 (``Trace the origin, growth and merger history of massive black holes across cosmic ages''), and in particular SI2.1 (``Search for seed black holes at cosmic dawn'') and SI2.3 (``Observation of EM counterparts to unveil the astrophysical environment around merging binaries of the LISA mission''):
\begin{itemize}
    \item With respect to SI2.1, the loss of $M<10^3\msun$ sources at $z>10$ caused by gaps is substantial. For the popIII model investigated, the detection rate of such sources is reduced from $\approx5$ yr$^{-1}$ for continuous observation streams ($T_{\rm elapsed}=T_{\rm data}$)  to $\approx 2$ yr$^{-1}$ in the case of the observations with gaps and a duty cycle of 75\%. Numbers can be as low as $\approx 1$ yr$^{-1}$ for more pessimistic scenarios. It is therefore clear that configurations T4G5 and T4G1 imply a significant loss of detections compared to the SciRD LISA baseline. The only way to mitigate the effect of gaps is by extending the mission duration.
    Therefore, in order to collect a large enough sample of such sources to ascertain the origin of seed MBHs, an extension to a 6-yr mission requirement (i.e.\ cases T6G5 and T6G1) is warranted.
    \item With respect to SI2.3, the detection rate of $M>10^5\msun$ sources at $z<2$ is of the order of $2$~yr$^{-1}$ in the investigated models. Because of gaps, about $30\%$ of them will suffer a significant loss of SNR compared to continuous collection of data throughout the mission lifetime, making parameter estimation and, particularly, sky localization problematic. In light of these considerations and in order to maximize the multi-messenger potential of MBHBs, an extension to a 6-yr mission requirement is warranted.
\end{itemize}
Conversely, gaps have a minor impact on SI2.2 (``Study the growth mechanism of MBHs before the epoch of reionization'') and SI2.4 (``Test the existence of intermediate-mass black holes''), as they do not pose a critical risk to the detection of the sources relevant for achieving those scientific goals.

\section{Stellar-mass compact objects}
\label{sec:WP2}

In this section we will study the impact of mission duration on resolved and unresolved stellar-mass sources (Sec.~\ref{subsec:stellar}) and on the observability of stellar-origin black holes (SOBHs) similar to those detected by the LIGO Scientific \& Virgo Collaboration (Sec.~\ref{subsec:SOBHs}).

\subsection{Stellar-mass sources}
\label{subsec:stellar}

Maximizing the number of detectable binaries is important to reduce the level of the confusion noise, which further improves the detectability and measurement accuracy of extra-Galactic sources at those same frequencies.  This is true even of transients which might occur during the first years of observations, as the improved understanding of the Galactic foreground can be applied retroactively when reanalyzing data from early in the mission.

\subsubsection{Resolved Sources}

Most of the resolved Galactic and extra-Galactic sources at low frequency will be nearly monochromatic, with evolution times much greater than both $\Tobs$ or $\Telapse$. Thus, gaps will not have strong effects on the majority of the resolved Galactic sources. However, in the cases where the frequency evolution occurs on similar timescales to the duty cycle, e.g., SOBHs (see Sec.~\ref{subsec:SOBHs}), gaps can reduce the fidelity of the parameter estimation of these sources.

\begin{figure*}
    \centering
    \includegraphics[width=\textwidth]{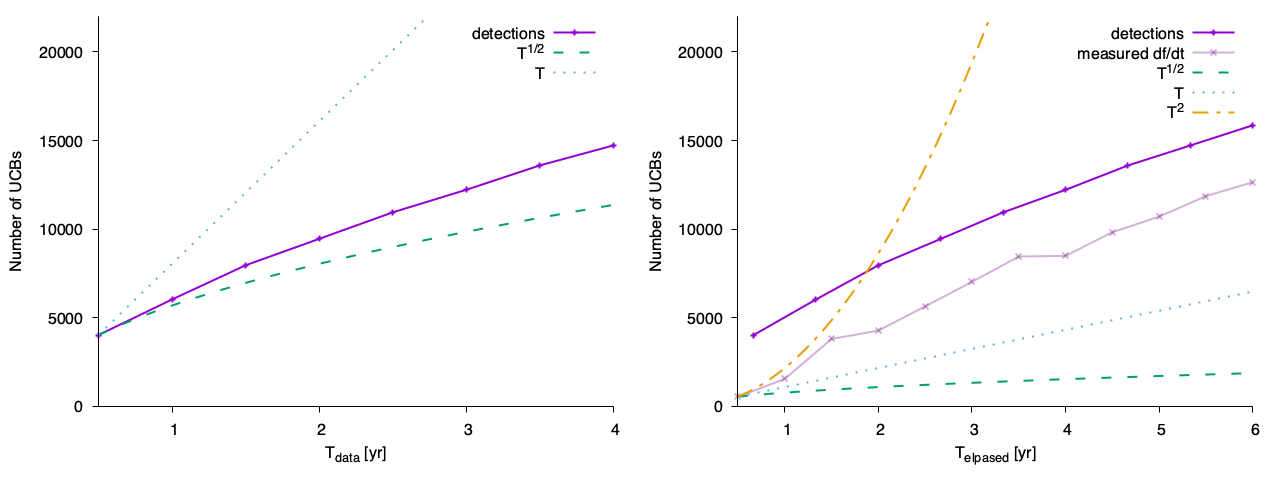}
    \caption{Left: The number of detectable UCBs scales between $\sqrt{\Tobs}$ and $\Tobs$ due to the combined effects of the increased SNR and frequency resolution. Right: The number of detected binaries with measurable $\dot f$ (used for breaking degeneracy between chirp mass and luminosity distance, and for identifying interacting binaries) scales more dramatically with \emph{elapsed} time $\Telapse$, because it enters the GW phase as $\Telapse^2$.}
    \label{fig:ucb_v_T}
\end{figure*}

The Galactic binary signals qualitatively scale as 
\begin{align}
h\propto \sqrt{\Tobs} \exp\left[i\pi\left(f\Telapse + \frac{1}{2}\dot{f}\Telapse^2 + \ldots\right)\right].
\end{align}
For an isolated binary the SNR scales as $\propto \sqrt{\Tobs}$ regardless of duty cycle when not considering losses of data due to windowing or TDI interpolation kernels. Therefore, longer observations are better, but the growth slows down as the observing time increases: the number of resolved Galactic binaries will increase much more quickly between years 1 and 2 of observing than between years 5 to 6.
However, in the confusion-dominated regime, the ability to distinguish resolvable binaries from the foreground depends on improved frequency resolution, which scales as $\propto 1/\Tobs$. As a result, the number of detectable binaries grows more rapidly than the simple SNR scaling predicts. The actual number of detections lands somewhere in the middle between $\sqrt{\Tobs}$ and $\Tobs$ (see Fig.~\ref{fig:ucb_v_T}, left panel).

Detailed studies of the Galactic binary population, and the dynamics of individual binaries, depend on measuring the time derivatives of the orbital period. These time derivatives introduce stronger time dependence, but importantly, it is the \emph{elapsed time} that matters most. The first time derivative of the frequency $\dot f$ is used to distinguish between systems that are likely evolving primarily due to GW emission vs.\ astrophysical interactions (e.g., mass transfer~\cite{2004MNRAS.349..181N, 2017ApJ...846...95K, 2018ApJ...854L...1B}). In cases where the orbital evolution is dominated by GW emission, $\dot f$ can break degeneracies in the GW amplitude to determine the sources' chirp mass and luminosity distance. Ref.~\cite{2019MNRAS.490.5888L, 2020MNRAS.494L..75S} show that the characterization of  $\dot f$ with mission durations of $4$ and $8$~yr leads to an increase from $\sim1100$ to $\sim2800$ double white dwarfs (DWDs) and $4$ to $10$ binary black holes (BBHs) with measured masses.

The $\dot f$ contribution to the GW phase scales as $\Telapse^2$, thus at fixed $\Tobs$ the science requirements for Galactic binaries benefit from lower duty cycles (see Fig.~\ref{fig:ucb_v_T}, right panel).
The second derivative of frequency depends even more dramatically on observing time, scaling as $\Telapse^3$. The second derivative of the orbital period encodes further details about dynamics (e.g., tidal interactions between binaries!\cite{2012ApJ...745..137V}) and gives an independent measure of chirp mass as a consistency test in the case of assumed GR-dominated period evolution. Systems with measurable $\ddot f$ will be comparatively rare, with $\mathcal{O}(10)$ sources providing constraints to better than $\sim20\%$ after $\Telapse \sim 8$~yr.

While a longer observing time from a longer mission duration will yield more resolved sources, in the case where duty cycles are being considered, maximizing $\Telapse$ is more impactful to SI1.1 (``Elucidate the formation and evolution of Galactic binaries
by measuring their period, spatial and mass
distributions'') and SI1.2 (``Enable joint gravitational and electromagnetic observations of Galactic binaries to study the interplay between gravitational radiation and tidal dissipation in interacting stellar systems''), than maximizing $T_{\rm{data}}$ alone.

\subsubsection{Unresolved foreground}
The unresolved foreground confusion noise can be characterized as~\cite{Karnesis:2021tsh}
\begin{equation}
S_{\mathrm{gal}} = \frac{A}{2} e^{-(f/f_1)^\alpha} f^{-7/3} \left[ 1 + \mathrm{tanh}\left( \frac{f_\mathrm{knee} - f}{f_2} \right) \right]\,
\label{eq:galfit}
\end{equation}
where $f$ is the frequency, $f_1$ and $f_2$ are the break frequencies, $f_\mathrm{knee}$ is the knee, $A$ is the overall amplitude, and $\alpha$ is a smoothing parameter.

This reduced empirical model was adopted after performing the analysis described above in this section, on the same catalog of sources, but considering different durations of the mission. Based on simulated LISA TDI time series data with total observation duration of 
$T_\mathrm{data,\, max} = 10$~yr, and estimated confusion noise for different fractions of $T_\mathrm{data,\, max}$, the parameters $f_1$ and $f_\mathrm{knee}$ of Eq.~\eqref{eq:galfit} are related to the observation duration $T_{\mathrm{data}}$ as:
\begin{eqnarray}
\begin{array}{r@{}l}
\log_{10}\left( f_1 \right) &= a_1 \log_{10} (T_{\mathrm{data}}) + b_1, \\
\log_{10}\left( f_{\mathrm{knee}} \right) &= a_k \log_{10}(T_{\mathrm{data}}) + b_k,
\end{array}
\label{eq:galfittobs}
\end{eqnarray}
where the parameters $a_1$, $a_k$, $b_1$, and $b_k$ depend on the SNR threshold for detectability of Galactic binaries. One of the most relevant characteristics of this unresolved foreground is $f_{\rm{knee}}$, which roughly indicates the boundary between the stochastic and resolvable parts of the signal and scales as $f_{\rm{knee}} \sim T_{\rm{data}}^{-0.4}$, a rather mild function of the observation time. However, the reduction in the stochastic foreground has an important impact on the SNR of other sources.

\subsection{Stellar-origin black holes (SOBHs)}
\label{subsec:SOBHs}

For the purpose of investigating the impact of mission duration on SOBHs, we considered: $T_{\rm data}=T_{\rm elapsed}=\{3,\,3.75,\,4.5\}$~yr of continuous data (T4C, T5C, T6C); and $T_{\rm elapsed}=\{4,\,5,\,6\}$~yr with a duty cycle ${\cal D}=0.75$ (T4G1/T4G5, T6G1/T6G5).

SOBHs generally have observable signal durations such that $T_{\rm{signal}} > T_{\rm elapsed}$. This makes the assessment of the impact of mission duration less straightforward compared to, e.g., MBHBs. The signal duration $T_{\rm signal}$ is also much longer than the duration of typical gaps thus, to first order, gaps will simply cause the SNR of the source to diminish by ${\cal D}^{1/2}$. To simulate the impact of data with gaps, we therefore artificially reduce the amplitude of the GW signal by ${\cal D}^{1/2}$, where ${\cal D}=0.75$; because of this, configuration T4C is essentially equivalent to T4G1/T4G5, and configuration T6C is equivalent to T6G1/T6G5. 

\begin{figure*}
    \centering
    \includegraphics[width=0.75\textwidth]{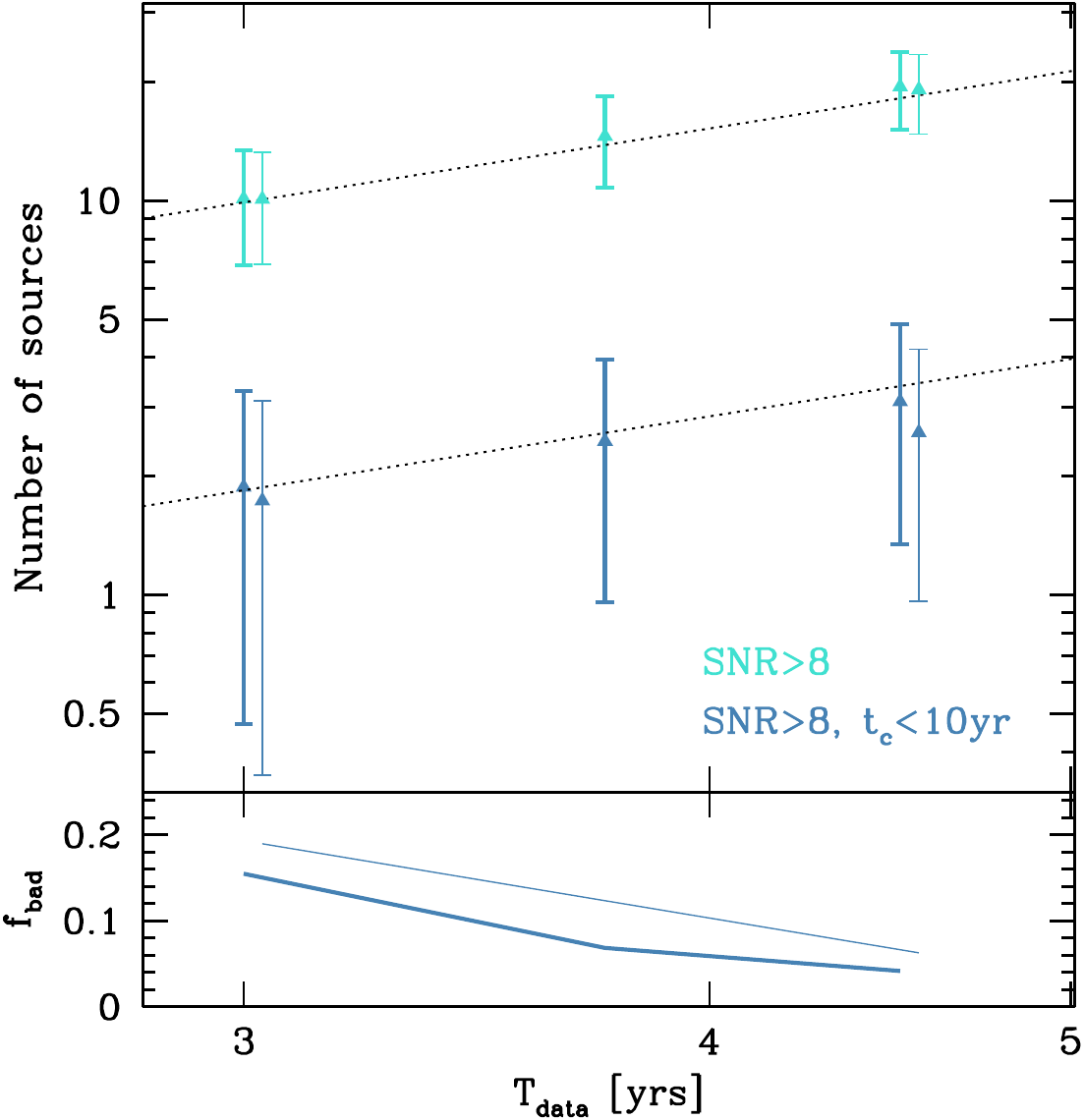}
    \caption{Upper plot: average number (with standard deviation) of detected SOBHs as a function of $T_{\rm data}$. All sources with SNR$>8$ are shown by the cyan marks, and the multiband subset is shown by the dark-blue marks, as indicated in figure.
      Thick marks are for continuous data (T4C, T5C, T6C) and thin marks for data with gaps and a duty cycle of 75\% (T4G1/T4G5, T6G1/T6G5). Thin marks have been slightly shifted to the left (by 0.005~dex) for readability. Black dotted lines show $N\propto T_{\rm data}^{3/2}$ for comparison. Lower plot: fraction of SOBH realizations resulting in no multiband source detected with SNR$>8$ as a function of time. Thick and thin lines are for continuous data (T4C, T5C, T6C) and data with gaps (T4G1/T4G5, T6G1/T6G5), respectively.}
    \label{fig:SOBH}
\end{figure*}

To investigate the effects of changes in mission duration, a SOBH population was simulated with a comoving merger rate density of $35$~Gpc$^{-3}$~yr$^{-1}$, with masses distributed flat in log space and a maximum mass cut for the primary BH of $M_1=50\msun$. We show the results of 1000 realizations of LISA observations for two scenarios (continuous data or data with gaps) in Fig.~\ref{fig:SOBH}. We find that the number of SOBHs that can be identified with SNR$>8$ increases from an average of 10, for 3 years of continuous data to an average of 19, for 4.5 years of data. This corresponds to a $N\propto T_{\rm{data}}^{3/2}$ scaling. The number of SOBHs observed by LISA depends on $T_{\rm data}$ rather than $T_{\rm elapsed}$. In practice, 4.5~yr of continuous observations yield the same number of detections as 6~yr of observations with 75\% duty cycle, since the gap duration of both the T6G1 and T6G5 scenarios are much shorter than $T_{\rm{signal}}$.
The number of potential multiband sources observable by ground-based detectors within 10 years of LISA observation also roughly doubles when increasing $T_{\rm data}$ by 50\% in scenarios T6C/T6G1/T6G5, going from $\approx 1.5$ to $\approx 3$, again assuming SNR$>8$.
By increasing $T_{\rm data}$ from 3~yr to 4.5~yr, the chance of a simulated Universe realization yielding zero multiband sources with SNR$>8$ ($f_{\rm bad}$, shown at the bottom of Fig.~\ref{fig:SOBH}) decreases from $\approx 20\%$ to $\approx 5\%$.

These findings have an impact on SO4 (``Understand the astrophysics of stellar origin black holes''), both SI4.1 (``Study the close environment of SOBHs by enabling multi-band and multi-messenger observations at the time of coalescence'') and SI4.2 (``Disentangle SOBH binary formation channels'') of the LISA proposal. The possibility of observing extra-Galactic SOBHs with LISA has been realized following the detection of GW150914. Early investigations suggested that LISA might observe up to several hundreds such sources, with few tens of them qualifying as multiband sources~\cite{2016PhRvL.116w1102S}. Subsequent downward revisions of SOBH merger rates, together with the relaxation of the LISA high-frequency sensitivity requirement, severely affected the expected numbers of SOBHs, jeopardizing the achievement of SOs listed in the LISA proposal. 

A 4~year mission with a 75\% duty cycle (T4C, T4G1, T4G5) will observe on average between 1 and 2 multiband sources with SNR $>8$, with a 20\% chance of observing none, thus completely failing the the SI4.1 science objective. Extending the mission requirement to 6~yr (T6C, T6G1, T6G5) will result in a rough doubling of multiband sources, reducing the risk of getting zero such sources to $\approx 5\%$.
Disentangling competitive SOBH formation channels based on eccentricity measurements for science objective SI4.2 requires a sizable number of detections. For example, based on calculations from Ref.~\cite{2017MNRAS.465.4375N}, the $\approx 10$ detections expected for $T_{\rm data}=3$~yr (T4C, T4G1, T4G5) will not even allow us to distinguish between the main field and cluster formation scenarios at a 2$\sigma$ level. Already with $\approx 20$ observations, allowed by $T_{\rm data}=4.5$~yr (T6C, T6G1, T6G5), the discriminating power will increase to $>3\sigma$. 

The detection numbers reported above are ultimately very sensitive to the intrinsic SOBH rate and to the maximum BH mass allowed by the pair instability gap. In particular, the existence of SOBHs with $M>50\msun$ would significantly increase the number of LISA detections. The SOBH landscape will become clearer with the release of the complete catalog of LIGO--Virgo O3 data. Given our current knowledge, extending the mission duration requirement to 6 years might be crucial to achieve SO4 of the LISA proposal.

\subsubsection{Detecting SOBHs from O1/O2 LIGO--Virgo catalogs} 

For concreteness, we consider the three loudest BBH systems in the LISA band from the O1/O2 LIGO--Virgo catalog~\cite{LIGOScientific:2018mvr}: GW150914, GW170104 and GW170823. For each of these three systems we find the best (for LISA) sky position and polarization. We estimate the SNR distribution based on posterior samples from the Gravitational Wave Open Science Center \cite{RICHABBOTT2021100658}, assuming that the system merges in 10~yr from the moment of observation.  By considering an observation time $T_{\rm data}$ and a 100\% duty cycle, we find the SNR values summarized in Table~\ref{SNR_O1O2}. In addition, given the distribution of SNR we give the probability (in percentage) of the source being above the detection threshold ($\rm{SNR} > 8$). As an example, for GW150914 optimally positioned on the sky, we find a best SNR of $12.34$ (for 6 years of observation),  a mean SNR of $7.21$ (based on the parameters uncertainties inferred by the LIGO--Virgo Collaboration), and a probability of having SNR $>8$ after 6~yr of observation of $\approx 25$\%.

\begin{table}[ht]
\caption{LISA SNRs for the three loudest systems from the O1/O2 LIGO--Virgo catalog: GW150914, GW170104 and GW170823.}
\label{SNR_O1O2}
 \begin{tabular}{|p{2.5cm}|p{2.5cm}|p{2.5cm}|p{2.5cm}|}
    \hline  
       & $T_{\rm{data}}$=4~yr & $T_{\rm{data}}$=5~yr & $T_{\rm{data}}$=6~yr\\ [0.5ex]
    \hline
      \multirow{3}{*}{GW Event}
      & max SNR & max SNR & max SNR\\ [0.5ex]
      & mean SNR & mean SNR & mean SNR\\ [0.5ex]
      & p(SNR $>$ 8) & p(SNR $>$ 8) & p(SNR $>$ 8)\\ [0.5ex]
    \hline\hline
 \multirow{3}{*}{GW150914} & 9.71 & 11.04 & 12.34  \\ 
   & 5.68 & 6.46 & 7.21 \\  
   & 2.27 & 12.21 & 25.50 \\
 \hline
 \multirow{3}{*}{GW170104} & 6.26 & 6.83 & 7.62  \\ 
   & 1.76 & 2.0 & 2.23 \\  
   & 0.0 & 0.0 & 0.0 \\
 \hline
  \multirow{3}{*}{GW170823} & 6.64 & 7.13 & 7.97  \\ 
   & 1.57 & 1.78 & 2.0 \\  
   & 0.0 & 0.0 & 0.0 \\
\hline
\end{tabular}
\end{table}

We now consider how parameter estimation for the three systems above is affected by the observation time. We vary the merger time between 7~yr and 20~yr from the start of LISA observation. Because these results are obtained using a Fisher matrix analysis, small fluctuations due to numerical evaluation of derivatives and inverting badly conditioned matrices are possible. 

\begin{figure*}
    \centering
    \includegraphics[width=1.0\textwidth]{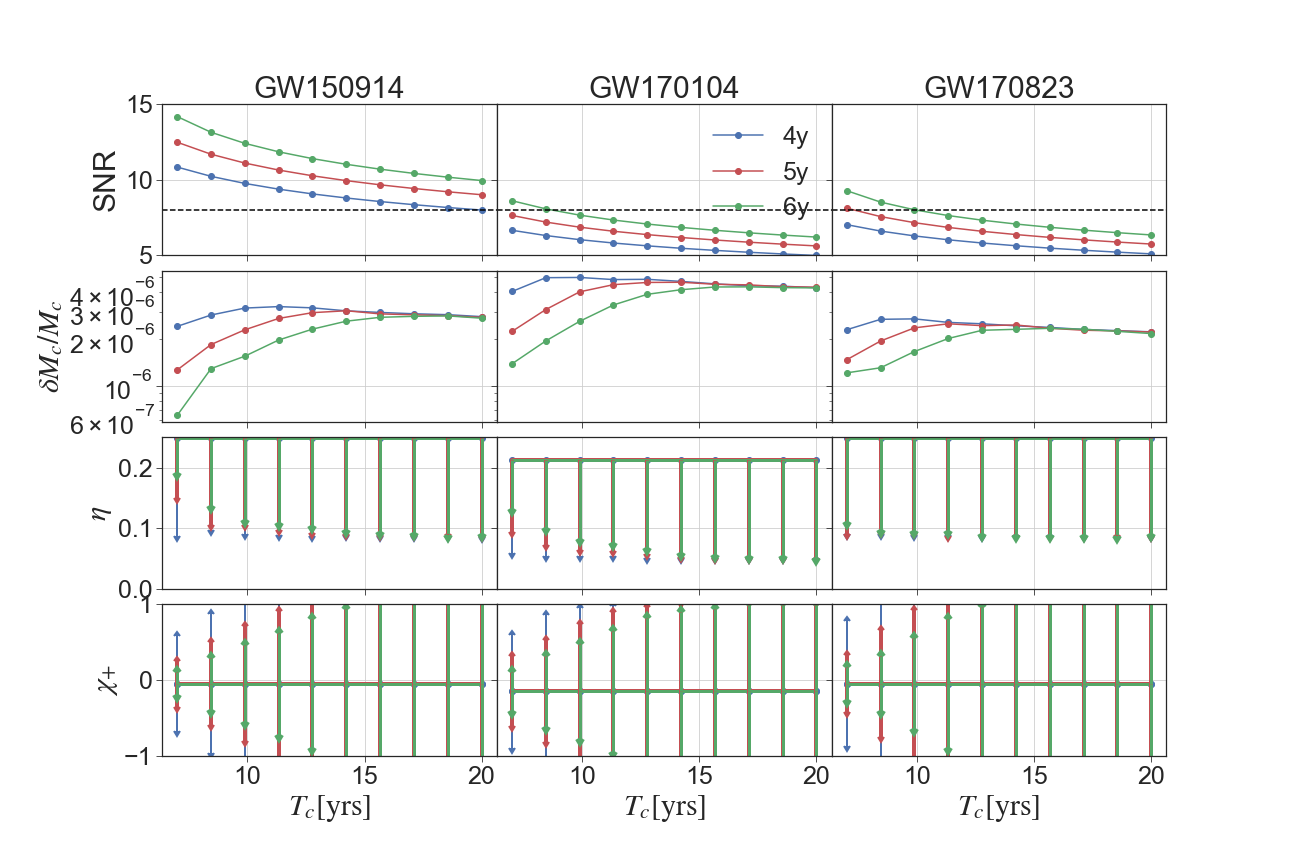}
    \caption{Parameter estimation evolution as a function of the time to coalescence $T_c$ and of the observational time.}
    \label{fig:LS}
\end{figure*}

For each source, in Fig.~\ref{fig:LS} we show the SNR, the relative error on the chirp mass $M_c$, and the absolute errors on the symmetric mass ratio $\eta$ and on the well-measured effective inspiral spin combination $\chi_+ = 
(m_1 \chi_1 +m_2\chi_2)/(m_1+m_2)$, where $\chi_1$ is the spin aligned with the orbital angular momentum of the primary, and $\chi_2$ is for the secondary.
The chirp mass is always well determined, while the mass ratio and spins are well determined only for systems which are not far from merging. With 4~yr of observation we can hardly constrain the mass ratio and spins, whereas 
with 6~yr of observation we can constrain the parameters of chirping systems.
The black dashed line corresponds to our (optimistic) detection threshold SNR = 8. Because these results are obtained assuming 100\% duty cycle, a lack of data from smaller duty cycles affects the SNR roughly as the square root of the duty cycle, so all reported errors will increase in the same proportion. 

The above results have direct impact on the detectability of GW150914-like systems as defined by science requirement SI4.1 and on evaluation of binary parameters for disentangling competitive SOBH formation channels defined by science requirement SI4.2. They support the recommendation that an extension of the mission lifetime to 6 years (T6C, T6G1, T6G5) is desirable.

\section{Extreme- and intermediate-mass ratio inspirals: detection, characterization, population}
\label{sec:WP3}

Extreme mass-ratio inspirals (EMRIs) consist of a stellar-mass compact object inspiralling into a MBH. 
The mass ratio is typically expected to be $\sim10^{-4}$--$10^{-6}$, meaning that the system completes many orbits emitting GWs in LISA's frequency band. 
Tracking the orbital evolution hence enables precision measurements of the system's properties and a characterization of the spacetime of the MBH. 
For this reason EMRIs are important for understanding the astrophysics of MBHs and their environments and for testing the Kerr nature of black holes. 
More extreme mass-ratio systems, such as those composed of a substellar-mass brown dwarf and massive black hole, are known as extremely large mass-ratio inspirals (XMRIs). 
These evolve even slower than EMRIS, negligibly changing over the lifetime of the LISA mission. 
Less extreme mass-ratio systems, such as either an intermediate-mass black hole and a MBH, or a stellar-mass compact object and an intermediate-mass black hole, are known as intermediate mass-ratio inspirals (IMRIs). 
These evolve quicker than EMRIs, and are more comparable to MBHBs or SOBHBs. 
We concentrate here on canonical EMRIs.

Changes to observing time, mission duration and gaps can effect the measured SNR (Sec.~\ref{sec:EMRI-SNR}), make it more difficult to track the phase (Sec.~\ref{sec:EMRI-plus}), and affect the total phase across the observations (Sec.~\ref{sec:EMRI-plus}). 
These effects can change the number of detections and the precision to which we can perform measurements.

\subsection{Changes in SNR}\label{sec:EMRI-SNR}

EMRIs are long-lived signals that accumulate their SNR over the observable lifetime of the inspiral. The number of detectable events increases faster than linearly with observing time $T_{\rm data}$. 
This is because while the number of EMRIs merging goes linearly with time, we also integrate for longer, meaning that quieter signals can accumulate sufficient SNR to become detectable.

The presence of gaps will decrease the SNR: to first order, the presence of a gap is effectively equivalent to changing the mission lifetime. The final parts of the signal are the loudest, so gaps during these times have the greatest cost.

\begin{figure*}
    \centering
    \includegraphics[width=0.70\textwidth]{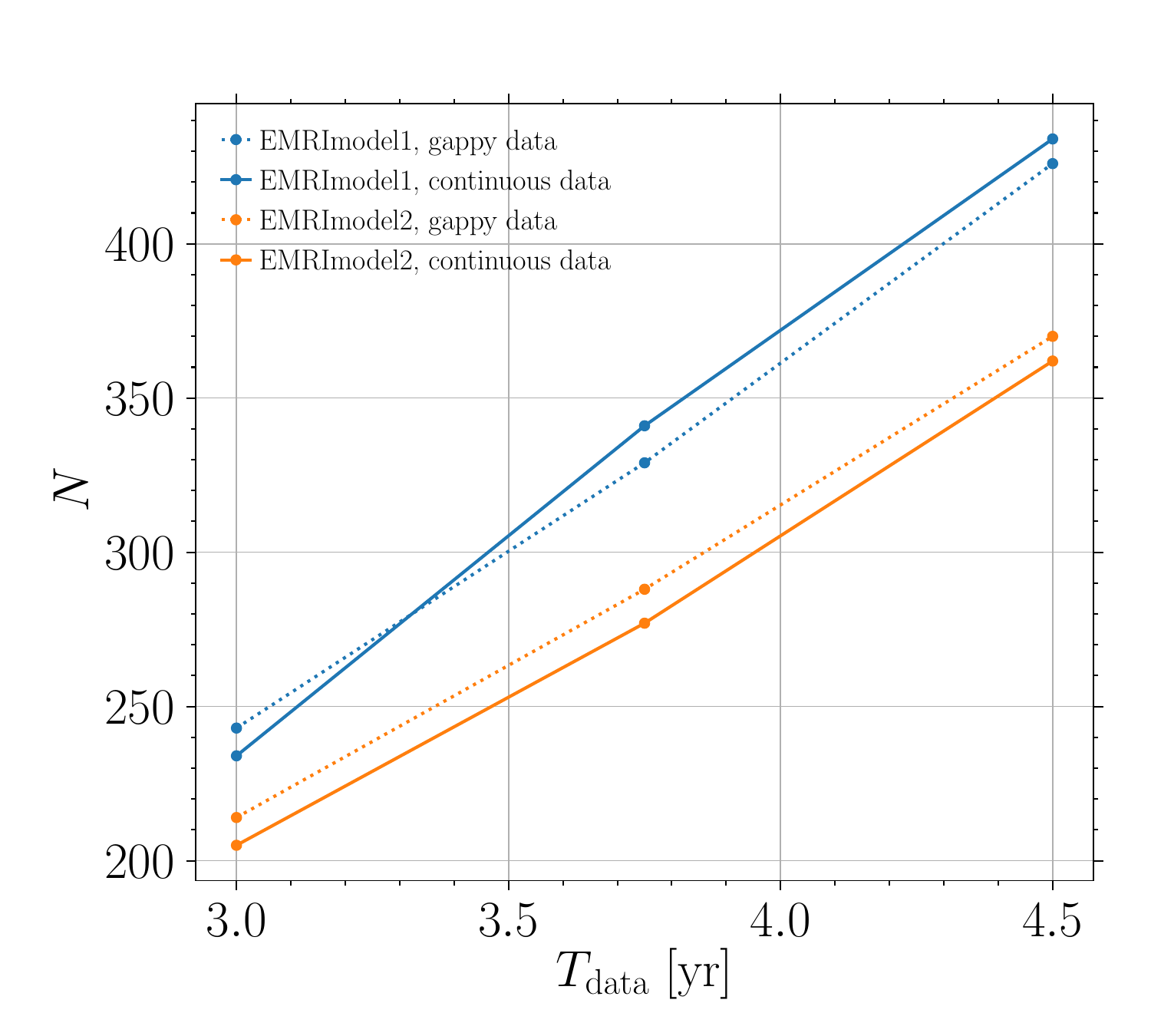}
    \caption{Number of EMRIs observed with SNR$>8$ as a function of $T_{\rm data}$ for two representative models from~\cite{2017PhRvD..95j3012B}. The plot shows that $T_{\rm data}$ sets the number of detections, regardless of the presence of gaps, and that the number of detections is roughly $\propto T_{\rm data}^{3/2}$.}
    \label{fig:EMRI}
\end{figure*}

To support these statements, we ran representative models from Ref.~\cite{2017PhRvD..95j3012B} with the same assumptions made for SOBHs in Sec.~\ref{sec:WP2}. 
Results are shown in Fig.~\ref{fig:EMRI}.
Similarly to the SOBH case, the number of observations is set by $T_{\rm data}$, regardless of the presence of gaps, and we find $N\propto T_{\rm data}^{3/2}$. 

Maximizing the potential for detection is extremely important if EMRIs are rare. 
This could be the case if tightly bound low-mass objects like brown dwarfs around MBHs are common~\cite{Gourgoulhon:2019iyu,Amaro-Seoane:2019umn}. 
These XMRI systems would not be detectable at cosmological distances, but they could disrupt the evolution of EMRIs, leading to scattering of the EMRI compact object before it enters the LISA band.
    
\subsection{Missing phase}\label{sec:EMRI-minus}

As EMRIs are long-lasting and slowly evolving signals, we should be able to track the GW phase across interruptions, enabling us still to perform matched filtering to dig the signals out of the noise. Complications arise if there is a more sudden and distinctive change in phase during a gap. 

A significant change in the phase evolution could happen if the EMRI passes through a transient resonance. 
These can occur due to radiation reaction in completely isolated systems (self-force resonance~\cite{Hinderer:2008dm}), or the tidal perturbation from a small third body (tidal resonance~\cite{Bonga:2019ycj}). 
Transient resonances are common, but only a few should have a noticeable impact~\cite{Ruangsri:2013hra,Berry:2016bit,Gupta:2021cno}. 
While missing the observation of a transient resonance would mean that we would not have the data at the time of the phase jump, this need not be a significant problem for detection or parameter estimation. 

Even though the change in phase is extremely sensitive to the orbital parameters on resonance, templates that account for resonance will still allow coherent filtering of the pre- and post-resonance data. 
This could be done in a fully modeled and self-consistent way~\cite{Gair:2010iv,Flanagan:2012kg}, or through the addition of phenomenological resonance parameters~\cite{Speri:2021psr}. 
An alternative approach is semicoherent analysis, which could enable the phase jump to be reconstructed without the use of resonance models.

\subsection{Extra phase}\label{sec:EMRI-plus}

Extending the mission lifetime $T_{\rm elapsed}$ means that there is potentially a greater observable phase change across the observing window. 
Assuming that the evolution can be tracked across the entire mission (even if semicoherent methods are used for initial detection, it may be possible to perform a coherent follow-up analysis), we can measure the total phase evolution, tracking its change with time even if there are gaps. 

The extended baseline gives greater sensitivity to quantities which affect the phase. This means greater measurement precision for parameters at a given SNR, which are essential for meaningful tests of relativity and the Kerr solution if the number of observed EMRIs is low. 
Measurements of environmental effects may also benefit from this extra observation time, as the phase change can increase superlinearly: for EMRIs in accretion disks, the scaling may be $\sim T_{\rm elapsed}^2$--$T_{\rm elapsed}^4$~\cite{Kocsis:2011dr}. 

Overall, since EMRIs are long-lived signals, data gaps are unlikely to cause a significant loss in scientific performance for astrophysics, provided that waveforms and analysis algorithms are developed to account for gaps. 
However, the presence of gaps will reduce the overall observing time, which could have an impact on the measurement precision. 
Long gaps might also discard valuable information about transient effects such as resonances, or potential high-frequency effects such as quasinormal bursts~\cite{Nasipak:2019hxh,Thornburg:2019ukt}.
An increase in mission lifetime enables observation of a greater change in phase, enabling more precise measurements at a given SNR, assuming that the phase can be tracked coherently across the entire duration.

In summary, although LISA's SO3 (``Probe the dynamics of dense nuclear clusters using EMRIs'') can likely be achieved by a 4-yr mission, several aspects of EMRI observations have superlinear scaling, indicating a clear preference for an extension of the mission lifetime requirement to 6~yr.

\section{Estimation of cosmological parameters}
\label{sec:WP4}

We report here on the impact of data stream duration with and without the presence of gaps on SI6.1 (``Measure the dimensionless Hubble parameter by means of GW observations only'') and SI6.2 (``Constrain cosmological parameters through joint GW and EM observations''). 

\subsection{Measurement of the Hubble parameter with EMRIs}

\begin{figure}
\centering
\includegraphics[width=8.5cm, height=5cm]{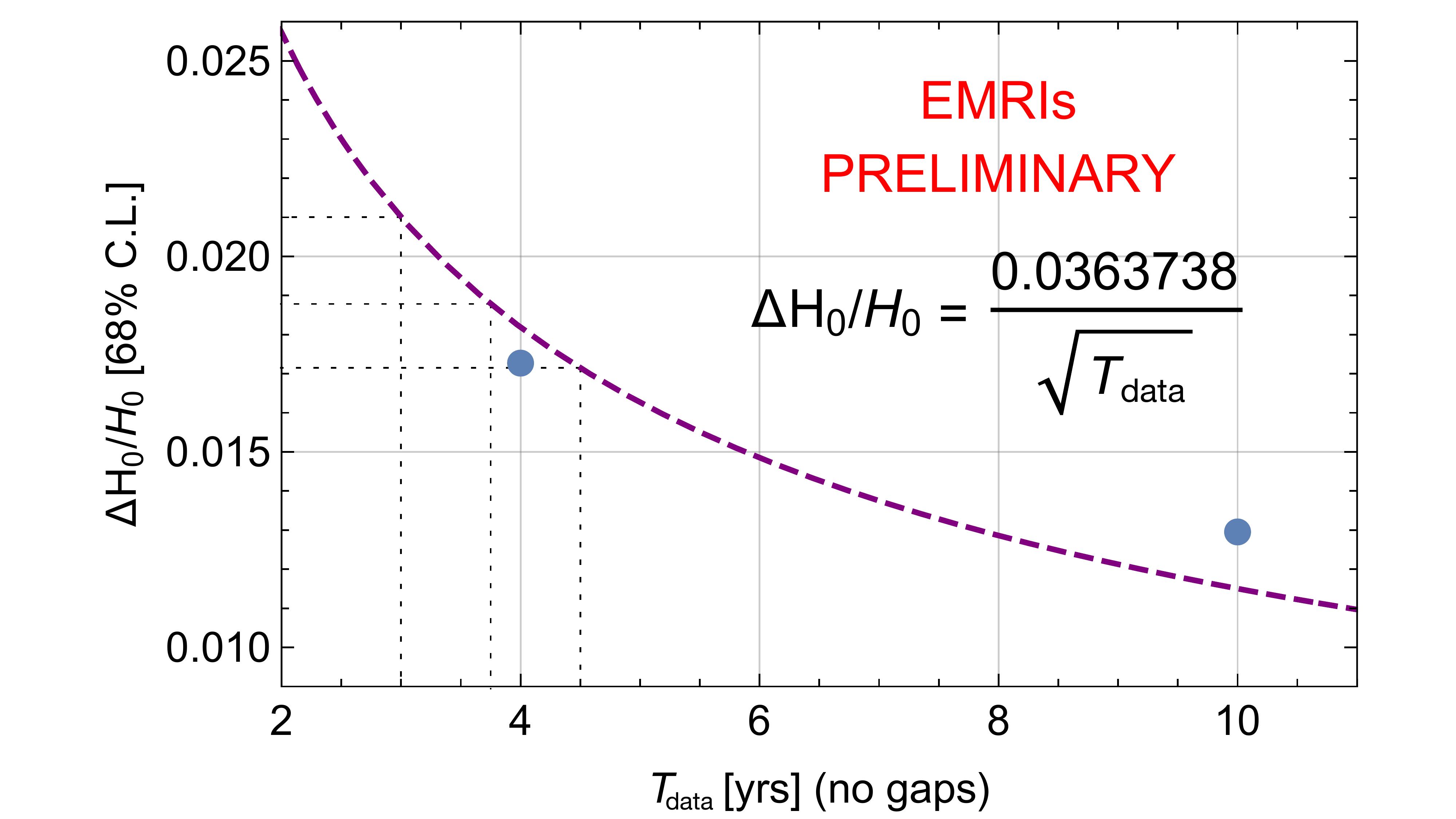}
\caption{Relative error on the Hubble parameter today, as a function of $T_{\rm data}$, from a preliminary analysis using EMRIs as distance indicators. The blue points show the two results obtained for 4 years of continuous data stream and 10 years of continuous data stream. The purple, dashed line shows the scaling proportional to $1/\sqrt{T_{\rm data}}$ that has been used to extrapolate the results for different data stream duration (3~yr, 3.75~yr and 4.5~yr, as shown by the dotted grey lines.)
}
\label{fig:EMRIsH0}
\end{figure}

In the SciRD, SI6.1 concerns the capability of LISA to constrain the Hubble parameter today, $H_0$, by using SOBHB and EMRIs as luminosity distance indicators, together with a statistical technique to identify the redshift, based on the cross-correlation of the GW measurement with galaxy catalogs. Preliminary results using only EMRIs as distance indicators hinted to the fact that with 4 years of continuous data it is possible to constrain the Hubble parameter today to about 1.7\% at 1$\sigma$ (cf.\ Fig.~\ref{fig:EMRIsH0}).
The analysis also considers 10~yr of continuous data, finding in that case the 1$\sigma$ uncertainty $\Delta H_0/H_0\simeq 1.3\%$. These results have since been confirmed by the analysis of Ref.~\cite{Laghi:2021pqk}, in the context of the most optimistic scenario for the EMRIs formation. 

Interpolating between these two results with a scaling of the relative error proportional to $1/\sqrt{T_{\rm data}}$ 
one would obtain that a 5-year mission with ${\cal D}=0.75$, corresponding to 3.75~yr of continuous data stream, is necessary to fulfill SI6.1, i.e.\ providing a measurement of $H_0$ to better than 2\% at 1$\sigma$.

\subsection{Measurement of the cosmological parameters with MBHBs}

We now turn to SI6.2, which refers to the capability of LISA to constrain cosmological parameters using MBHB as luminosity distance indicators, together with EM counterparts to determine the redshift.

For this analysis, we adopt the methodology developed in Ref.~\cite{Tamanini:2016zlh}. The technique to identify the counterpart can be either direct observation of the host galaxy (in particular, we modeled detection with the LSST), or connecting the GW source with a transient occurring at the moment of the MBHB merger, e.g., a radio jet. In this last case, we have implemented sky localization with the SKA and redshift identification from the host galaxy with the ELT~\cite{Tamanini:2016zlh}. We analyzed three astrophysical models for the formation of the MBH, two with high-mass seeds (Q3d, which provides the lowest number of sources, and Q3nd, which provides the highest one) and one with low-mass seeds (popIII, giving an intermediate number of sources)~\cite{Tamanini:2016zlh}.

\begin{figure}
    \centering
    \includegraphics[width=0.6\textwidth]{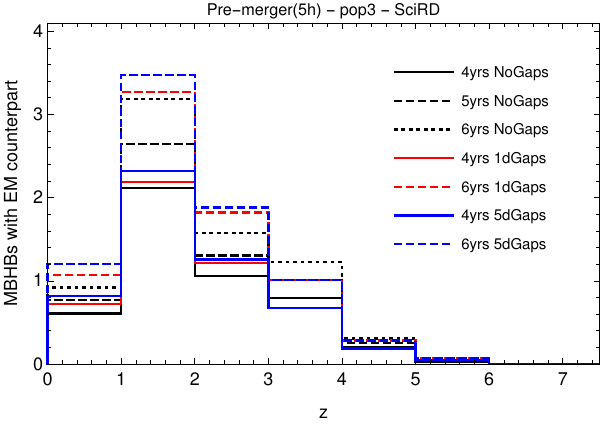}\\
    \includegraphics[width=0.6\textwidth]{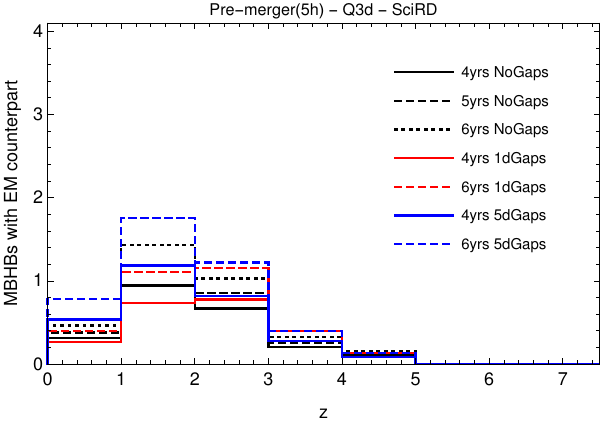}\\
    \includegraphics[width=0.6\textwidth]{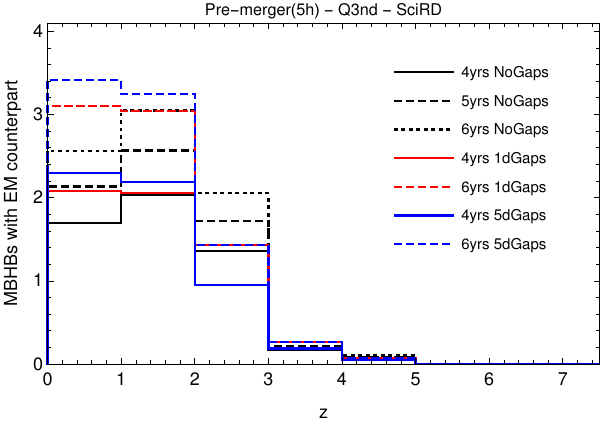}\\
    \caption{Number of standard sirens as a function of redshift for the 7 mission duration/gaps scenarios, from left to right in the low mass seed MBHB formation channel (popIII), and in the two high mass seeds ones (Q3d and Q3nd).}
    \label{fig:StSiz}
\end{figure}

We analyzed the following duration scenarios, all with ${\cal D}=0.75$: continuous data stream of 4~yr, 5~yr and 6~yr (T4C, T5C and T6C); 4~yr data stream with 1-day and 5-day gaps (T4G1, T4G5); and 6~yr data stream with 1-day and 5-day gaps (T6G1, T6G5).
Figure~\ref{fig:StSiz} shows the distribution of standard sirens as a function of redshift for the different duration scenarios. The majority of standard sirens resides in the redshift range $1<z<3$ (the more optimistic astrophysical model Q3nd presents a significant number of sources also at $z<1$). 
The number of standard sirens scales linearly with the data stream duration, and the scenario providing the highest number of standard sirens is T6G5. 
In scenarios with gaps, it is less likely to completely miss a source, while shorter and more frequent gaps lead to the highest SNR loss.  

\begin{figure}
    \centering
    \includegraphics[width=\textwidth]{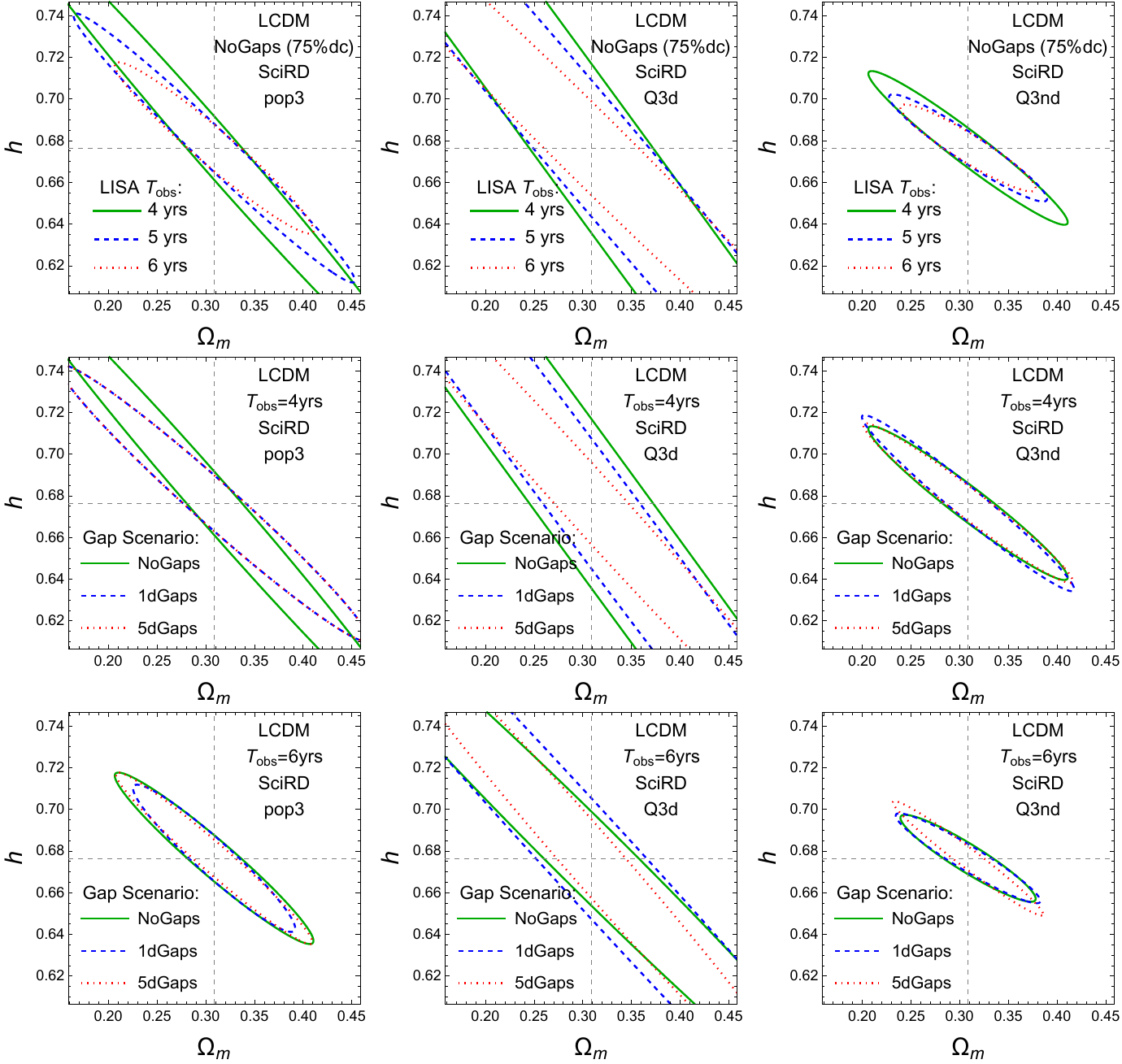}
    \caption{Estimated 1$\sigma$ ellipses of the relative uncertainties on the cosmological parameters, $h$ and $\Omega_m$, with 75\% duty cycle. Top row: continuous data stream of 4, 5 and 6~yr; middle row: 4~yr data stream with no gap, 1-day, and 5-day gaps; bottom row: 6~yr data stream with no gap, 1-day, and 5-day gaps. From left to right, different columns refer to different MBHB formation channels: popIII, Q3d, and Q3nd.}
    \label{fig:StSiell}
\end{figure}

In Fig.~\ref{fig:StSiell} and Table~\ref{tab:StSiFoM} we present the 1$\sigma$ relative uncertainties on $h$ and $\Omega_\mathrm{m}$, 
where $h = H_0 / (100 {\rm km\,s^{-1}\,Mpc^{-1})})$ and $\Omega_\mathrm{m}$ is the relative fraction of (dark) matter energy density today,
for all 3 MBHB astrophysical formation channels and all data stream duration scenarios,
The uncertainties naturally scale inversely to the square root of the number of standard sirens: therefore, the best case scenario is the one with 6~yr data stream, and 5-day gaps.

\begin{table}[ht]
\caption{Estimated 1$\sigma$ relative uncertainties on $H_0$ and $\Omega_\mathrm{m}$, for all 3 MBHB astrophysical formation channels~\cite{Tamanini:2016zlh} and all data stream duration scenarios.
}
\medskip
\centering
\includegraphics[width=\textwidth]{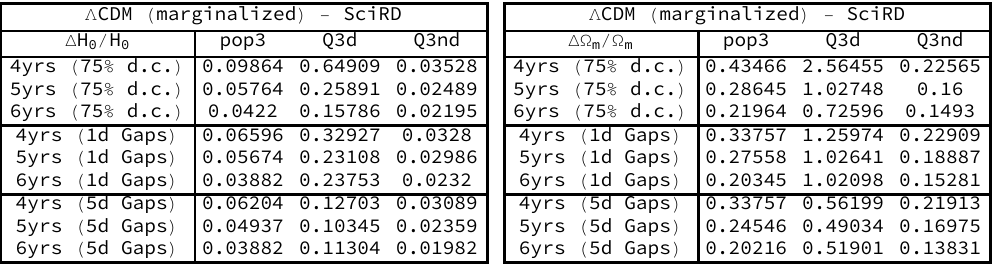}
\label{tab:StSiFoM}
\end{table}

We adopt as a Figure of Merit a threshold error on $H_0$ less than 4\% for at least two formation channels.  This is met by two of the duration scenarios: 6~yr data stream with 1-day and 5-day gaps, T6G1 and T6G5. 
The error on $H_0$ strongly depends on the MBHB formation channel. In the best case (Q3nd, featuring high-mass seeds with no delay in the binary formation) it is always smaller than $3.5\%$, while in the worst case (PopIII, featuring low-mass seeds) it can grow to as much as $65\%$ for the T4C mission configuration. 

As a consequence of  the lack of a full parameter-estimation analysis including merger and ringdown on the MBHB catalogs considered here~\cite{Tamanini:2016zlh}, the results presented above are based on the estimation of the MBHB event parameters performed accounting for the inspiral phase only (cut 5 hours before merger). This approach underestimates the number of available standard sirens, and consequently also the instrument performance. On the other hand, in the absence of catalogs produced with the SciRD sensitivity in the frequency range $10^{-4}~{\rm Hz}<f<0.1~\rm{Hz}$, we have used those produced with SciRD, but extended down to $10^{-5}$~Hz. This might overestimate the number of detected standard sirens, although measurements at low frequency are not expected to strongly affect the present analysis (cf.\ the low-frequency study~\cite{LISAlowfreq}). %

\section{Characterization of stochastic backgrounds}
\label{sec:WP5}

A stochastic GW background (SGWB) can be characterized by its power spectrum as a function of frequency, by the angular variation of its intensity~\cite{Bartolo:2016ami, Caprini:2018mtu}, and possibly by its polarization. 

The SNR for the measurement of an isotropic SGWB scales as $\sqrt{T_{\rm data}}$ 
under the assumption of stationary signal and noise \cite{Smith:2019wny}.
Therefore, the presence of randomly distributed 1-day or 5-day gaps influences the signal detection capability only as it influences the total duration of the data stream.\footnote{We work under the ideal assumption that gaps do not impact the quality of the data, while it is instead likely that the noise varies after a gap. Specific tools will be needed to address this issue.}
We thus analyze only the three scenarios without gaps: continuous data for 3~yr (Case T4C), continuous data for 3.75~yr (Case T5C), continuous data for 4.5~yr (Case T6C). We perform two kinds of studies (cf.\ the low-frequency study~\cite{LISAlowfreq}).
The first one, presented in Sec.~\ref{sec:PLSGWB}, concerns the generic power-law signal
\begin{equation}
    \Omega_{\rm GW}(f)= \Omega_0 (f/1\, {\rm mHz})^{n_\mathrm{T}}, \label{eq:powerlaw}
\end{equation}
and the specific signals defined in SI7.1 (``Characterise the astrophysical stochastic GW background'') and SI7.2 (``Measure, or set upper limits on, the spectral shape of the cosmological stochastic GW background'') of the SciRD \cite{SciRD}, which read
\begin{eqnarray}
{\rm SI~7.1:}~&&\Omega_{\rm GW}(f)=4.0\cdot 10^{-10}\left(\frac{f}{25\,{\rm Hz}}\right)^{2/3}  \theta(f- f_2 ) \,\theta(f_5 -f) ,
\label{eq:SciRD71}\\
{\rm SI~7.2:}~&&\Omega_{\rm GW}(f)=2.8\cdot 10^{-11}\left(\frac{f}{10^{-4}\,{\rm Hz}}\right)^{-1} \theta(f- f_1 ) \,\theta(f_4 -f) \nonumber
\\ && \qquad \qquad + \,8.0 \cdot 10^{-12}\left(\frac{f}{10^{-2}\,{\rm Hz}}\right)^{3}  \theta(f - f_3)\theta(f_6 -f),\label{eq:SciRD72b}
\end{eqnarray}
where $\theta (f)$ is the Heaviside step function and $f_{1,2,3,4,5,6}=\{0.1, 0.8, 2, 15, 20, 100\}$~mHz.
In Eq.~\eqref{eq:powerlaw}, $n_T$ is the primordial spectral index; this case is sufficiently general to describe a spectrum arising from inflation, scaling sources like cosmic strings, or the tail of a broken power-law as arising from a first order phase transition.
The spectrum given in Eq.~\eqref{eq:SciRD71} represents an astrophysical foreground of inspiraling binaries, characterized by the $f^{2/3}$ spectrum \cite{Phinney:2001di}.
Finally, to probe a broken power-law SGWB from the early universe, Eq.~\eqref{eq:SciRD72b} is a statement of the requisite sensitivity to achieve the target science goals \cite{2017arXiv170200786A}: it represents the minimal sensitivity requirement to detect either the infrared tail $f^3$ (if the peak is above $0.1$~Hz), or the ultraviolet tail $1/f$ (if the peak is below $0.1$~mHz), of a broken power-law signal from bubble collision during a first-order phase transition~\cite{Caprini:2015zlo}. This particular source has been chosen as a representative example. 
The second study, presented in Sec.~\ref{sec:EUSGWB}, considers the signals caused by two possible SGWB sources operating in the early universe.
Both studies show that changing the overall mission duration from 4~yr to 5~yr or 6~yr (i.e.\ 3~yr, 3.75~yr and 4.5~yr of continuous data stream) provides an insignificant detection improvement. In particular, SI7.1 and SI7.2 can be fulfilled in all three duration scenarios.

LISA is also sensitive to the angular variation of the SGWB intensity, as it has different sensitivity to different regions of the sky while orbiting around the Sun. 
The SNR for the detection of an SGWB anisotropy scales proportionally to $\sqrt{T_{\rm data}}$~\cite{Seto:2004np}. 
On the other hand, gaps could influence the SGWB anisotropy characterization, as they might reduce the detector sensitivity to a particular region of the sky. If they appear with a random pattern (i.e.\ at random positions of the LISA orbit),
it is conceivable that their influence is similar to the one of a reduction in the overall mission duration. The worst case scenario would be the one of gaps with periodicity multiple to one year, 
so that LISA would be always blind at the times in which it is mostly sensitive to a specific region of the sky. However, we can foresee that LISA will be able to pick up the anisotropy of the SGWB only at very large scales, represented by the first few multipoles of the spherical harmonics expansion of the sky, say $\ell \lesssim 10$. Gaps with duration of the order of a few days would correspond to a sensitivity loss at much smaller scales, for which the resolution of the instrument is already very low. On the basis of these arguments, we infer that the overall continuous data-stream duration, and the presence of gaps in the data stream, do not significantly alter the capability of LISA to characterize the anisotropy of the SWGB.

\subsection{Analysis of power law SGWB signals}
\label{sec:PLSGWB}

\begin{figure}
\includegraphics[width=0.9\linewidth]{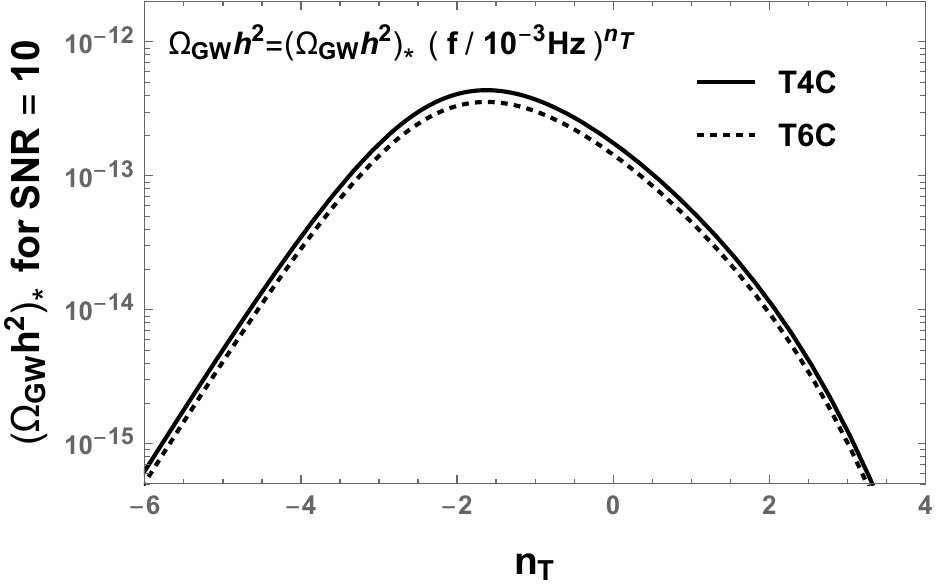} 
\caption{Contour regions of parameter space in which the SGWB signal $\Omega_{\rm GW}(f) = \Omega_0 (f/1~{\rm mHz})^{n_T}$ has SNR $>10$. This has been calculated using the SciRD sensitivity curve, for 3~yr (solid line) and 4.5~yr (dashed line) of continuous data stream.}
\label{fig:SGWBSNR}
\end{figure}

\begin{figure}
\includegraphics[width=0.9\linewidth]{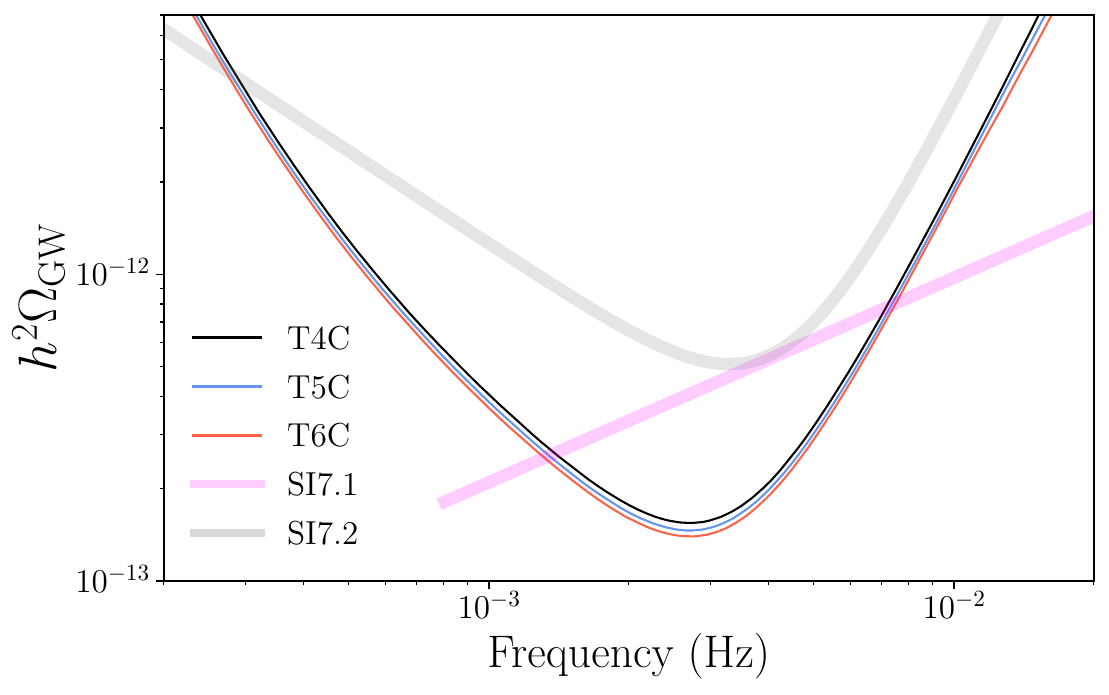}
\caption{The coloured contour lines represent the level of the signal amplitude $\Omega_{\rm GW}$ that would be detected with high confidence, $\mathcal{B}\geq 100$, for the three continuous data stream duration scenarios. The grey and pink lines represent the signals identified in SI7.1 and SI7.2, given in Eq.~\eqref{eq:SciRD71})--Eq.~\eqref{eq:SciRD72b}.}
\label{fig:SGWBBayes}
\end{figure}

To quantify the effect of increasing the overall continuous data-stream duration on the SGWB detection, we analyse the detection capabilities for the signals in Eq.~\eqref{eq:powerlaw}, Eq.~\eqref{eq:SciRD71} and Eq.~\eqref{eq:SciRD72b} for the duration scenarios T4C, T5C and  T6C. For the signal in Eq.~\eqref{eq:powerlaw}, we adopt the fiducial detection criterion $\textrm{SNR}\!>\!10$. For the duration scenarios T4C and T6C, this criterion is fulfilled in the parameter region $\{\Omega_0,n_\mathrm{T}\}$ below the solid curve and the dashed curve of Fig.~\ref{fig:SGWBSNR}, respectively. The result highlights that for this kind of signal, the gain in parameter space from T4C to T6C is too small to justify an extension of the mission duration. We further investigate the detectability of the SGWBs in Eq.~\eqref{eq:SciRD71} and Eq.~\eqref{eq:SciRD72b}, %
with a more elaborated detection criterion.  
Specifically, we adopt the Bayes factor $\mathcal{B}$
between a model with pure noise and a model with noise plus a generic power-law SGWB signal (see Ref.~\cite{Karnesis:2019mph} for details). The result is shown in Fig.~\ref{fig:SGWBBayes}: the signals of SI7.1 and SI7.2 given in Eq.~\eqref{eq:SciRD71}--Eq.~\eqref{eq:SciRD72b} do satisfy $\mathcal{B}\geq 100$, meaning that they can be detected with high confidence also in the shortest mission duration Case T4C. Besides being detected, these signals are also reasonably well reconstructed. We use the \texttt{SGWBinner} code~\cite{Caprini:2019pxz,Flauger:2020qyi} to test this feature.

\begin{figure}
\centering \includegraphics[width=0.9\linewidth]{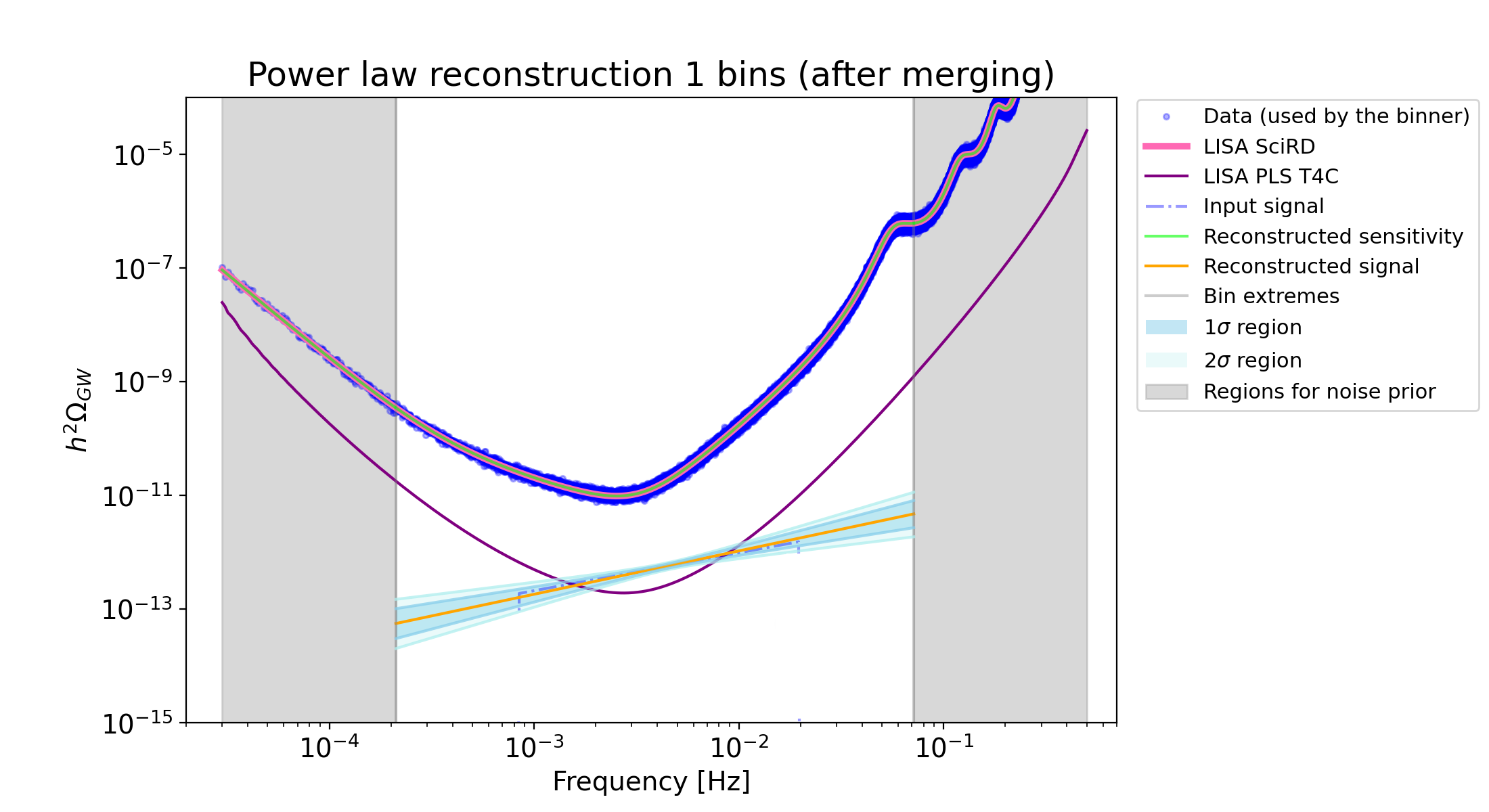} 
\caption{Reconstruction of the SI7.1 signals by the \texttt{SGWBinner} code for the mission duration scenario T4C. The algorithm converges to a 1-bin reconstructionm with the bin being  approximately $[2\times 10^{-4}~\textrm{Hz}, 7\times 10^{-2}~\textrm{Hz}]$. The remaining frequency region is used to improve the prior on the noise parameters.}
\label{fig:SGWBREC-7.1}
\end{figure}

\begin{figure}
\centering 
\includegraphics[width=0.9\linewidth]{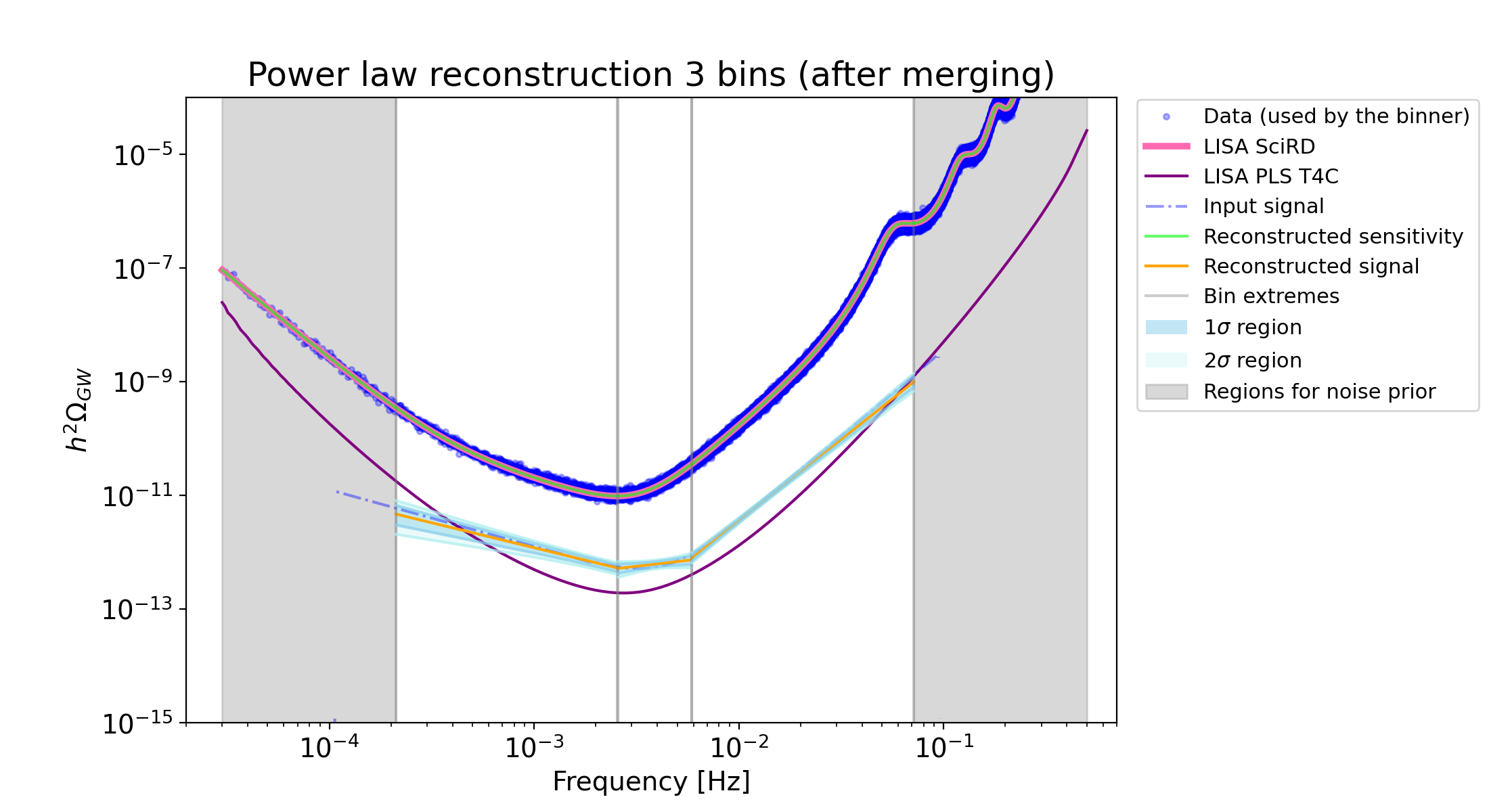}
\caption{Reconstruction of the SI7.2 signal by the \texttt{SGWBinner} code for the mission duration scenario T4C. The algorithm converges to a 3-bin reconstruction with the bins being approximately $[2\times 10^{-4}, 3\times 10^{-3}]~Hz$, $[3\times 10^{-3}, 6\times 10^{-3}]$~Hz and $[6\times 10^{-3}, 7\times 10^{-2}]$~Hz. The remaining frequency region is used to improve the prior on the noise parameters.}
\label{fig:SGWBREC-7.2}
\end{figure}

\begin{figure}
  \centering
\includegraphics[width=0.8\linewidth]{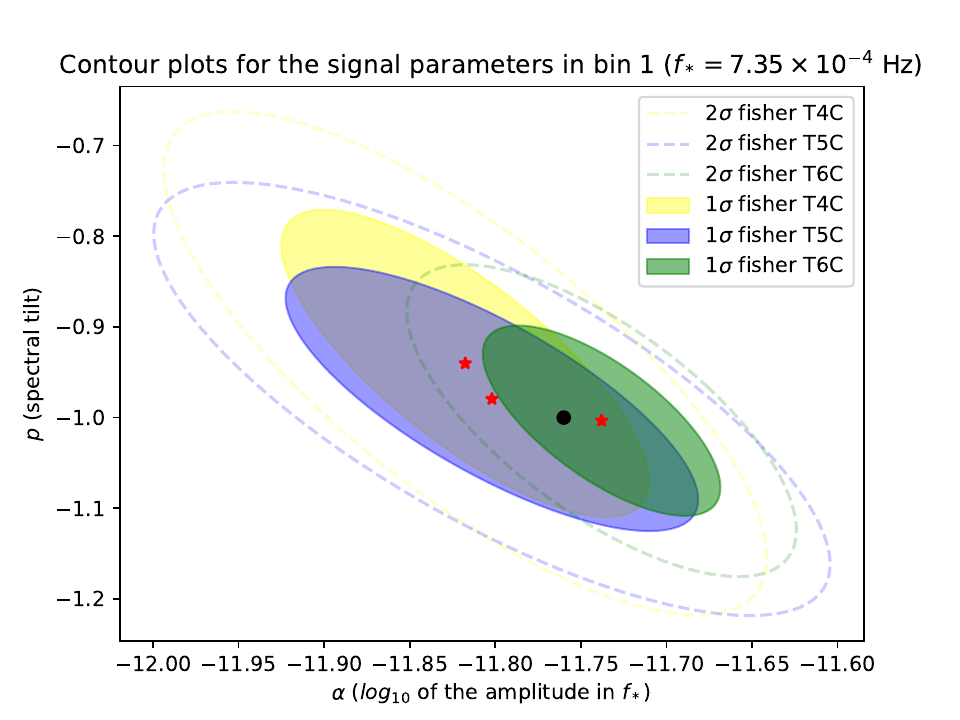} 
\includegraphics[width=0.8\linewidth]{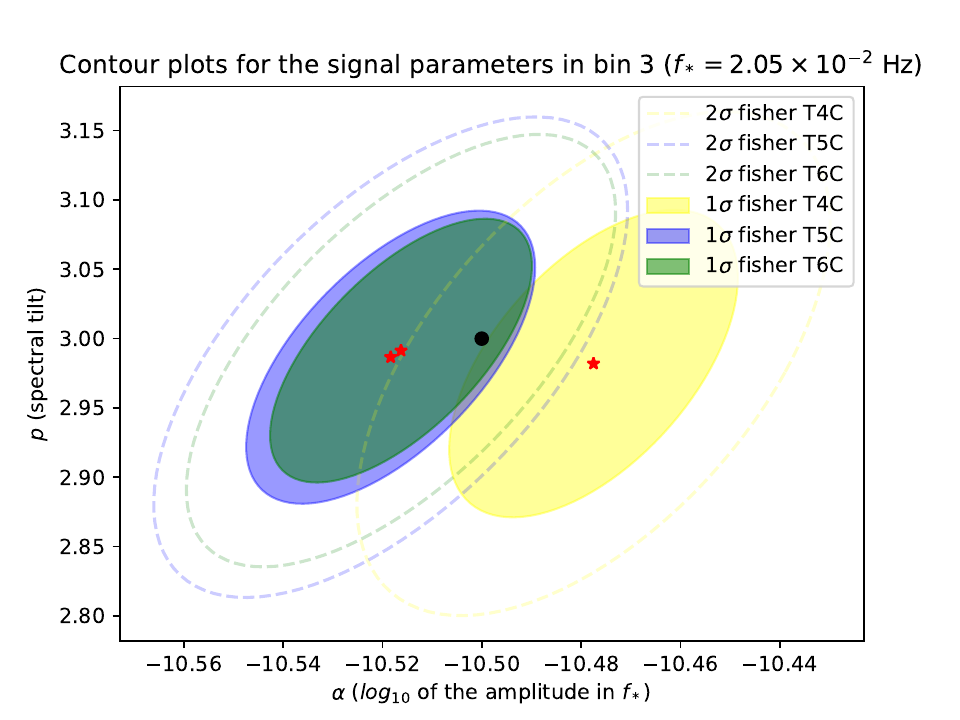}
\caption{1$\sigma$ (shaded area) and 2$\sigma$ (dashes lines) Fisher ellipses of the SI7.2 signal reconstruction via the \texttt{SGWBinner} reconstruction shown in Fig.~\ref{fig:SGWBREC-7.2}. The left and right panels shows to the ellipses reconstruction in the outermost left and right bins in Fig.~\ref{fig:SGWBREC-7.2}. The reconstructed parameters $\{\Omega_0,n\}$ are evaluated in term of the pivot frequency $f_*$ indicated in each panel. The black dots represent the true parameter values, while the red ones represent the reconstructed best fit values. They are all compatible within 1$\sigma$.}
\label{fig:SGWBBin3el}
\end{figure}

The \texttt{SGWBinner} code reconstructs the spectral shape of a SGWB signal in the LISA band, via parameter estimation of a series of power laws fitting the signal in frequency bins with adaptive size (the noise curve parameters are also  reconstructed at the same time). In each bin, the reconstruction follows the parametrization $\Omega_{\textrm GW}= \Omega_0 (f/f_*)^n$. 
At this stage of code development~\cite{Flauger:2020qyi}, we use a single TDI channel~\cite{Caprini:2019pxz} as the consequent reconstruction improvements would rely on extra assumptions on the LISA noise.
%
Figures~\ref{fig:SGWBREC-7.1} and \ref{fig:SGWBREC-7.2} display the reconstruction perspective in the duration scenario T4C in the case of the SI7.1 and SI7.2 signals, respectively. Both signals  can be reconstructed with reasonably small error bars even in the shortest mission duration scenario T4C.
In particular, the left panel and right panel of Fig.~\ref{fig:SGWBBin3el} show the 1$\sigma$ and 2$\sigma$ Fisher ellipses of the reconstructed parameters for the signal SI7.2 in the left outermost and right outermost reconstruction bins, respectively. Different colors correspond to different duration scenarios. 
In all duration scenarios, the reconstructed parameters are compatible with the true values (black dots) within 1$\sigma$. 
In the cases T5C and T6C, the areas of the 1$\sigma$ ellipses are  $\sim$1.1 and $\sim$1.4 times smaller than the area in the case T4C. 
The areas scale approximately linearly with $T_{\rm data}$, corresponding to relative errors on the reconstruction parameter decreasing as $\sqrt{T_{\rm data}}$. The gain of 20\% in the parameter reconstruction of these signals is a target that should have lower priority than other possible improvements in the LISA mission.

\subsection{Analysis of early universe sources}
\label{sec:EUSGWB}

\begin{figure}
  \centering
\includegraphics[width=0.8\linewidth]{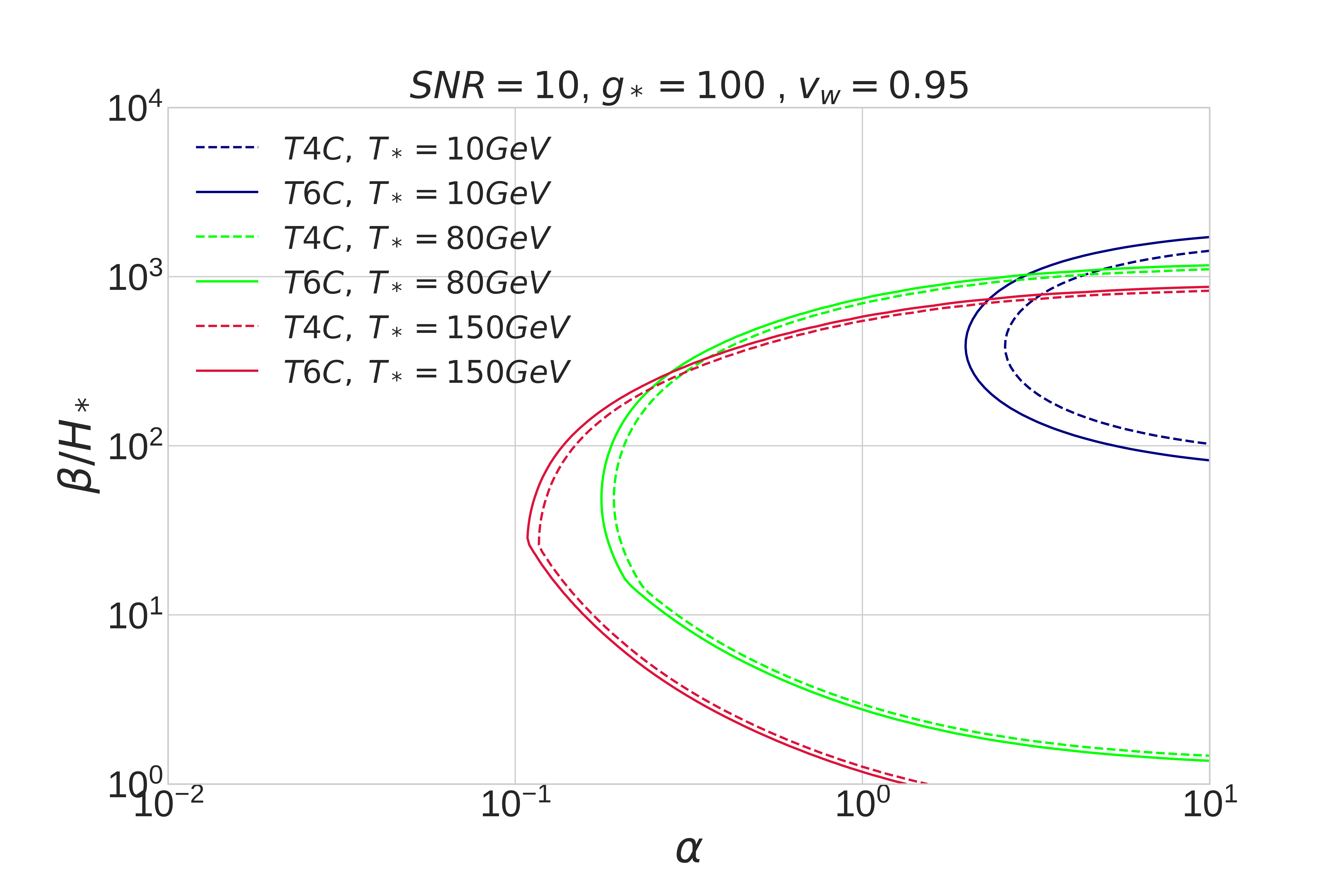} 
\caption{The parameter region $\{\alpha, \beta/H_* \}$ that LISA can probe when the FOPT SGWB is dominated by the sound-wave contribution. The regions on the right of the curves, evaluated for some given values of $v_w, g_*, T_*$, 
are detectable with ${\rm SNR}>10$. Solid lines correspond to the scenario T4C while the dashed ones to T6C.}
\label{fig:SGWBFOPT}
\end{figure}

A first order phase transition (FOPT) occurring in the primordial universe can generate a SGWB detectable by LISA. The FOPT parameters entering the SGWB signal are the transition temperature $T_*$, strength $\alpha$, inverse relative duration $\beta/H_*$ and the bubble wall velocity $v_w$. Several mechanism can source GWs: bubble wall collisions, and the thereby generated sound waves and/or magnetohydrodynamic turbulence~\cite{Caprini:2015zlo}. Here we focus on the GW signal produced by sound waves, the one that is best characterized~\cite{Caprini:2019egz}. Fixing the FOPT temperature to 10~GeV, 80~GeV and 150~GeV, the bubble wall velocity to a highly relativistic value $v_\mathrm{w}=0.95$, and the number of relativistic degrees of freedom to $g_*=100$, we quantify the gain in parameter space from increasing the continuous data stream duration. The result, shown in Fig.~\ref{fig:SGWBFOPT}, is that the extra parameter region reached by increasing the mission duration from 4~yr to 6~yr with ${\cal D}=0.75$ is too small to prioritize an extension of the mission (for details on the codes, see Refs.~\cite{ptplot, source_figs}). Concerning the 7442 FOPT benchmark points identified in \cite{Caprini:2019egz}, the variation in the detection prospects increases as: 478/7442 points for T4C; 516/7442 points for T5C; 538/7442 points for T6C.

\begin{figure}
  \centering
\includegraphics[width=0.8\linewidth]{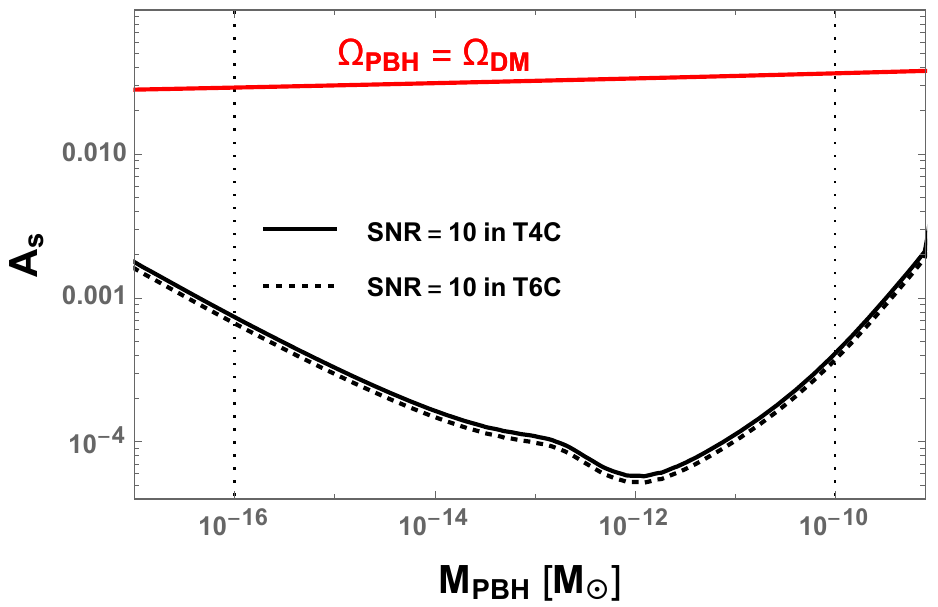}
\caption{Red curve: amplitude of the scalar power spectrum that gives the totality of the dark matter being PBH, as a function of the PBH mass. Black curves: minimal amplitude needed to have ${\rm SNR}=10$ at
LISA for T4C (solid line) and T6C (dotted line). The dotted vertical lines denote approximately the mass range of interest: the lower bound originates from the $\gamma$ background due to PBH evaporation, and the higher bound originates from lensing (Subaru HSC).}
\label{fig:SGWBPBH}
\end{figure}

A similar result is obtained in the case of the SGWB signal generated by second-order scalar perturbations, when these latter are enhanced by the presence of a bump in the primordial inflationary scalar power spectrum (see for instance \cite{DeLuca:2019qsy} for the details of the computation). Figure~\ref{fig:SGWBPBH} shows that, not only the gain in parameter space is tiny, but the range of the parameter space which is scientifically the most relevant is well within the reach of the three mission duration configurations. This corresponds to the range, in the amplitude of the bump of the scalar spectrum, for which this inflationary scenario leads to primordial black holes (PBHs) with masses that allow them to account for 100\% of the dark matter in the Universe.

\section{Constraints on dark matter}
\label{sec:WP6}

Many theoretical models predict the existence of ultralight boson fields, which may be a significant fraction of the dark matter content in the Universe.  Because of black hole superradiance, these fields may be sources of monochromatic GWs that can be detected either from isolated sources or as a stochastic background~\cite{Brito:2015oca}. The analysis of Ref.~\cite{Brito:2017zvb} indicates that extending the mission duration from 4~yr to 6~yr would increase LISA's sensitivity to resolvable and stochastic sources of this kind. The number of detectable events with phase coherent searches scales as $T_{\rm data}^{3/2}$, while semicoherent searches scale as $T_{\rm data}^{3/4}$.  For resolvable continuous GWs, this translates into a factor of ${\sim} 1.8$ (${\sim} 1.4$) increase in the number of sources detectable by a coherent (semicoherent) search.  Mission duration also impacts the boundaries in parameter space of the expected constraints on boson masses. By extending the analysis of Refs.~\cite{Brito:2017wnc,Brito:2017zvb} to a more general mission duration,
we find a difference in the interval of masses probed by this method of around $5{-}10 \%$ (e.g., a 4-yr mission would constrain dark matter with particle masses in the range $[3.7 \times 10^{-19}, 2.3\times 10^{-16}]$~eV, while a 6-yr mission would constrain the range $[3.3\times 10^{-19}, 2.7\times 10^{-16}]$~eV). However, these numbers are heavily dependent on astrophysical population models that have large uncertainties.\\

\begin{figure}
    \centering  \includegraphics[width=7.5cm]{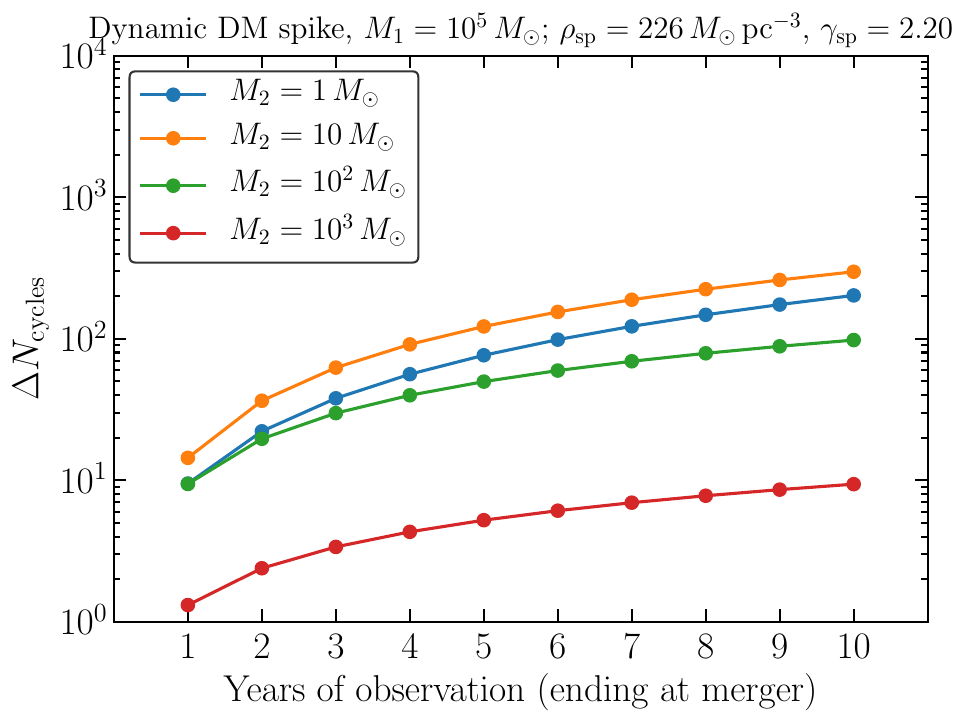}
\includegraphics[width=7.5cm]{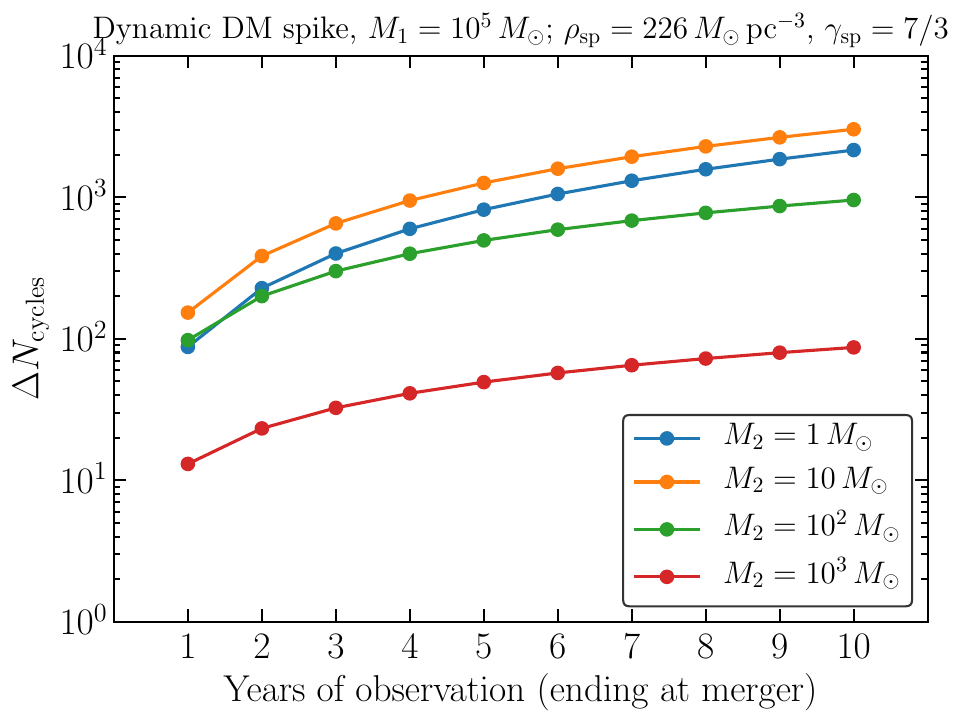}    \caption{Change in the number of cycles due to a DM spike with respect to the vacuum case, for different total observation times, with the observation ending at the merger. These results were obtained by adapting the code developed in Ref.~\cite{Kavanagh:2020cfn}. They refer to a central IMBH of mass $M_1 = 10^5 \msun$ and different masses $M_2$ for the smaller compact object, as shown in the legend. The difference between the two plots is in the properties of the DM spike (parametrized by $\gamma_{\rm sp}$). \label{fig:DMprofile}}
\end{figure}

Searches for dark matter imprints on gravitational waveforms are not as developed~\cite{Eda:2013gg,Eda:2014kra,Barausse:2014tra,Hannuksela:2018izj,Cardoso:2019rou,Hannuksela:2019vip,Kavanagh:2020cfn}.
Approaches using Newtonian expressions for dynamical friction, incorporating accretion but no backreaction on fluid-like dark matter configurations, find that the post-Newtonian (PN) phasing is affected at $-5.5$PN order~\cite{Barausse:2014tra,Cardoso:2019rou}. For models where dark matter is an ultralight field the correction is a $-6$PN effect~\cite{Annulli:2020lyc}. The impact of the duration of the mission can be estimated by connecting this phenomenology to the PN parameters (cf.\ Sec.~\ref{sec:WP7}). In most situations, the difference between a 4-yr and 6-year mission is a factor of 2 improvement in the constraints on dark matter density. This general prediction was confirmed by large $N$-body simulations of IMRIs in some particular scenarios~\cite{Kavanagh:2020cfn}. Figure~\ref{fig:DMprofile} shows the dephasing in the GW signal for two DM profiles, as a function of mission duration. The dephasing grows linearly (or faster) with the observation time. In some cases where the dephasing may be marginal (of order 1 cycle), increasing the observation time can be important for getting an effect large enough to be detectable.

\section{Tests of general relativity}
\label{sec:WP7}

We now ask how the LISA mission duration affects our ability to test general relativity (GR) with LISA. We quantify the effect of mission duration by using parametrized tests and inspiral--merger--ringdown consistency tests.

\subsection{Parametrized tests} 
In GR, the GW signal in the time domain
can be written in the form $h(t,k)=\mathcal{A}_\mathrm{GR}(t,k)e^{i \Phi_\mathrm{GR}(t,k)}$,
where $\mathcal{A}_\mathrm{GR}(t,k)$ is the amplitude and $\Phi_\mathrm{GR}(t,k)$ is the phase of the wave.
These two quantities are the main observables. 
Non-GR effects can be classified into two categories: emission effects and propagation effects. Emission and propagation effects can modify both the amplitude and the phase of GW signals~\cite{Yunes:2016jcc,Tahura:2018zuq,Nishizawa:2017nef}. 

Let us first discuss the non-GR corrections to the amplitude. 
The amplitude is given by an initial amplitude at emission $\mathcal{A}^i_\mathrm{GR}(t,k)$ multiplied by the transfer function $\mathcal{T}_\mathrm{GR}(t,k)$ encoding information about the cosmological evolution, i.e.\  $\mathcal{A}_\mathrm{GR}(t,k)=\mathcal{A}^i_\mathrm{GR}(t,k)\mathcal{T}_\mathrm{GR}(t,k)$. 
Corrections due to modified emission can be simply mimicked by taking the appropriate modified function $\mathcal{A}^i_\mathrm{non-GR}(t,k)$ as the initial condition. If the background evolution is not $\Lambda$CDM, one would capture that with an appropriate transfer function $\mathcal{T}_\mathrm{non-GR}(t,k)$. 
The precise measurement of the amplitude will be for instance crucial for the GW luminosity distance, enabling us to provide an independent measurement of the expansion rate $H_0$. Since there will be degeneracies between the dimming of the amplitude due to the expansion and due to new physics, one will need to theoretically model and observe the merger rate of compact binaries as a function of redshift. For instance, if the gravity theory contains additional non-abelian gauge fields~\cite{BeltranJimenez:2018ymu} or tensor fields~\cite{Max:2017flc} belonging to the dark sector, they will yield a periodic effect on the amplitude due to GW oscillations. These effects can be parametrized in a model-independent way and tested against the redshift information. Therefore, the LISA mission duration will be crucial to obtain good statistical rates to break such degeneracies~\cite{Jimenez:2019lrk}. In the following we will solely focus on the modifications in the waveform phase and work in the Fourier domain.

Non-GR corrections to the inspiral part of the waveform phase in the Fourier domain can be prescribed within the parametrized post-Einstein (ppE) formalism~\cite{Yunes:2009ke} (or generalized IMRPhenom formalism~\cite{TheLIGOScientific:2016src,Abbott:2020jks}, that has a one-to-one correspondence with the ppE parametrization for corrections entering in the inspiral waveform~\cite{Yunes:2016jcc}) as
\begin{equation}
\label{eq:PPE}
\Psi = \Psi_\mathrm{GR} + \beta u^{2n-5}\,,
\end{equation}
where $\Psi_\mathrm{GR}$ is the waveform phase in GR and $u \equiv (\pi \mathcal M f)^{1/3}$.\footnote{The effect of a non-$\Lambda$CDM cosmological evolution on the phase may be captured by introducing a transfer function similar to the amplitude case. We can effectively take such effects into account below if the corrections to the phase fall within the ppE parametrization.} Here $\mathcal{M}$ and $f$ denote the chirp mass of the binary and the GW frequency, $\beta$ represents the non-GR correction parameter, and the index $n$ indicates that the correction enters at $n$th PN order relative to GR. Such a theory-agnostic formalism can be mapped to violations of various fundamental aspects of GR, such as the strong equivalence principle (time variation of $G$ at $-4$PN, scalar dipole radiation at $-1$PN), Lorentz invariance ($-1$PN and 0PN), parity invariance (2PN), or a nonzero graviton mass (1PN)~\cite{Yunes:2016jcc,Tahura:2018zuq}. Such a formalism also allows us to probe dark matter effects (e.g., gravitational drag at $-5.5$PN or $-6$PN~\cite{Cardoso:2019rou,Annulli:2020lyc}) and frequency-dependent departures of the GW propagation speed from $c_\mathrm{T}=1$ (in this case, the PN order depends on the form of the dispersion relation).

\begin{figure}
    \centering \includegraphics[width=7.5cm]{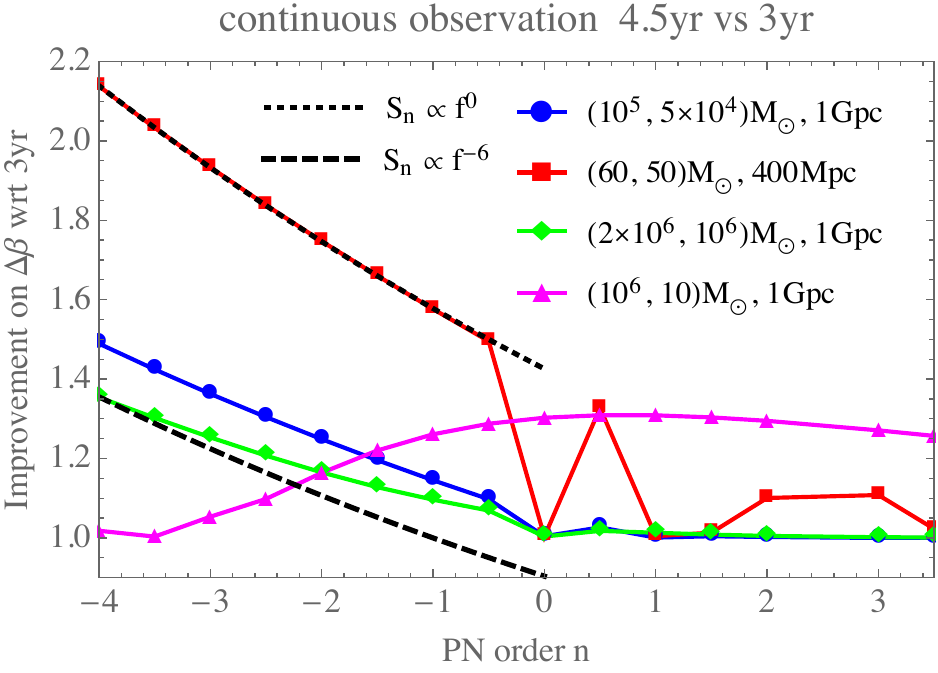}
\includegraphics[width=7.5cm]{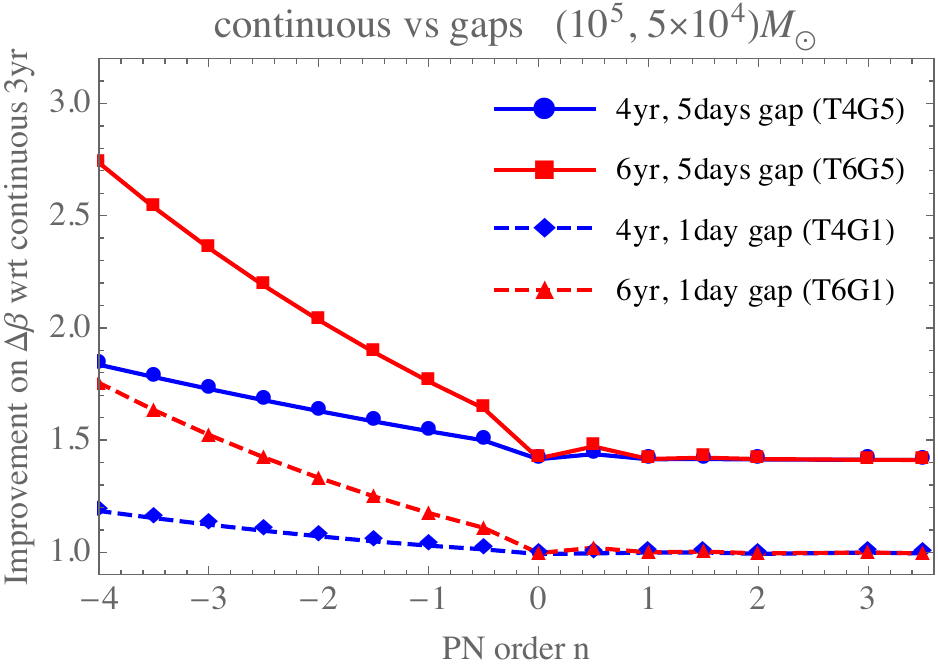}    \caption{Top: Improvement on constraining the non-GR parameter $\beta$ in the phase, cf.\ Eq.~\eqref{eq:PPE}, at different PN orders with a continuous $T_\mathrm{data}=4.5$~yr observation (scenario T6C in Sec.~\ref{sec:intro}) relative to a $T_\mathrm{data}=3$~yr observation (scenario T4C) for various example systems.  We assume that the observation starts at a time $T_\mathrm{data}$ before coalescence. The detector's low-frequency cutoff is assumed to be $10^{-4}$~Hz for all cases, except for the SMBH binary system $(2 \times 10^6,10^6)\msun$, for which we assumed the detector cutoff frequency to be at $10^{-5}$~Hz. If the cutoff frequency were at $10^{-4}$~Hz there would be {\em no difference} in terms of measuring $\beta$ between the 3~yr and 4.5~yr cases for this SMBH binary system (the frequency 3 years before coalescence is already outside of this cutoff frequency, and thus a longer observation time does not change the measurability of $\beta$).  We also show the rough analytic estimate of Eq.~\eqref{eq:beta_T_scaling}, or more precisely the quantity $(4.5/3)^{(4n-3s-14)/16}$ with $s=0$ and $s=-6$. Bottom: same as in the top panel, but now including gaps in the observation. We compare the measurability of $\beta$ for the 4 scenarios with gaps in Sec.~\ref{sec:intro} against the case with a continuous observation for 3~yr (T4C). We assumed that mergers occur outside of the gaps.}
    \label{fig:insp_tests}
\end{figure}

The top panel of Fig.~\ref{fig:insp_tests} presents the ratio of the upper bound on $\beta$ between continuous 3~yr vs.\ 4.5~yr observations. This ratio measures the improvement in tests of GR with 4.5~yr of observation relative to 3~yr of observation, and shows that the typical improvement is by a factor of 1--2. Following Ref.~\cite{Yunes:2016jcc}, the IMRPhenomD waveform has been used for the GR part of the waveform, and the measurability of $\beta$ is estimated through a Fisher matrix analysis. EMRIs have a different behavior from other systems, probably because the dynamical frequency range is small, and longer observations help to break the degeneracy between $\beta$ (at positive PN orders) and other parameters, like the masses. We assumed that the observation starts $T_\mathrm{data}$ before coalescence, which is the optimal case. If we cannot detect the merger, it would be difficult to break the degeneracy between $\beta$ and other parameters even for probing negative PN effects, and thus the measurability of $\beta$ becomes much worse than the case considered here.

The bottom panel of Fig.~\ref{fig:insp_tests} shows a similar result, but including gaps in observations. The bounds on $\beta$ can improve by a factor of 3 compared to the continuous 3~yr observation case. With a fixed elapsed time of 4~yr, the improvement is up to a factor of 2. This is because the case with gaps can have a wider dynamical frequency range when performing a Fisher analysis. We also see that longer gap durations yield better improvements at probing non-GR effects in these examples. This is possibly because there is a significant difference in the frequency evolution in the last segment of observation (that contains the merger) compared to all the other segments. The amount of frequency change for the case of 5-day gaps (with 15 days observation segment) is larger than for 1-day gaps (with 3 days observation segment), which further helps to break the degeneracy between $\beta$ and other parameters.

We can give a rough estimate of how $\Delta \beta$ scales with the observation time $T_\mathrm{data}$ at negative PN orders (at positive PN orders $\beta$ has strong correlations with other parameters, and thus it is not easy to find such a scaling). If we neglect correlations between $\beta$ and other parameters, $\Delta \beta$ is roughly given by
\begin{equation}
\Delta \beta \sim \frac{1}{\sqrt{\Gamma_{\beta\beta}}}, \qquad \Gamma_{\beta\beta} \equiv 4 \int_{f_\mathrm{min}}^{f_\mathrm{max}} \mathrm{d}f \frac{\partial_\beta \tilde h \partial_\beta \tilde h^*}{S_\mathrm{n}}\,.
\end{equation}
Here $f_{\min}$ and $f_{\max}$ are the minimum and maximum cut-off frequencies, $\tilde h$ is the waveform in Fourier space, and $S_n$ is the noise spectral density.
The absolute value of the waveform amplitude in frequency domain scales like $|\tilde h| \propto f^{-7/6}$ and $\partial_\beta \tilde h \equiv \partial \tilde h / \partial \beta \propto \tilde h  f^{\frac{2n-5}{3}}$. Assuming a simple scaling for the noise as $S_\mathrm{n} \propto f^s$, one finds
\begin{equation}
\Gamma_{\beta\beta}\propto \int_{f_\mathrm{min}} \mathrm{d}f \frac{[f^{-7/6} f^{(2n-5)/{3}}]^2}{f^{s}} 
\sim f_\mathrm{min}^{(4n-3s-14)/{3}}. 
\end{equation}
Assuming that we start the observation a time $T_\mathrm{data}$ before coalescence, we have $f_{\min} \propto T_\mathrm{data}^{-3/8}$, and therefore\footnote{The $f_{\min}$ dependence in Eq.~\eqref{eq:beta_T_scaling} agrees with Eq.~(33) of Ref.~\cite{Perkins:2020tra} when $s=0$, after accounting for the $f_{\min}$ dependence in the SNR.} 
\begin{equation}
\label{eq:beta_T_scaling}
\Delta \beta \propto f_{\min}^{-(4n-3s-14)/{6}} \propto T_\mathrm{data}^{(4n-3s-14)/{16}}\,. 
\end{equation}
We show this scaling in the top panel of Fig.~\ref{fig:insp_tests} for $s=0$ and $s=-6$. Observe that this analytic estimate with $s=0$ agrees almost perfectly with the numerical result for the system with $(60,50)M_\odot$. This is because $f_{\min}$ for such a system is $f_{\min} \sim 0.01$~Hz where $S_\mathrm{n} \propto f^0$. On the other hand, for systems with larger masses, $f_{\min}$ is much lower and the numerical results can be better captured with $S_\mathrm{n} \propto f^{-6}$, which is the frequency dependence of the noise at low frequency.
The deviation from this scaling is due to the various approximations used in this rough estimate, and in particular to the degeneracy between $\beta$ and other parameters. 

\subsection{Inspiral--merger--ringdown consistency tests} 

Another model-independent test of GR with GWs is the inspiral--merger--ringdown consistency test~\cite{Ghosh:2016qgn,Ghosh:2017gfp,TheLIGOScientific:2016src,Abbott:2020jks,Carson:2019kkh,Carson:2020cqb}, where we measure the final mass and spin of the remnant black hole with inspiral and merger-ringdown independently and check the consistency between the two measurements. We studied how such tests are affected by the mission duration for the two sources with masses $(10^5,5\times 10^4)\msun$ and $(60,50)\msun$ considered in the top panel of Fig.~\ref{fig:insp_tests}. 
As expected, the mission duration only changes the final mass and spin estimate from the inspiral portion, though the difference is small. We conclude that, at least for the systems studied here, the inspiral-merger-ringdown consistency tests are almost unaffected by the duration of the observation.

In summary, longer observation times mainly improve bounds on non-GR effects entering at negative PN orders (such as varying-$G$ effects) by a factor of 2--3. As shown in Table~\ref{tab:my_label}, it also helps to have more events, and hence better statistics.

\section{Testing the nature of black holes}
\label{sec:WP8}

\begin{table}[]
    \centering
    \scriptsize
    \begin{tabular}{||c|ccc|c||}
    \hline
    \hline
      {\bf Effect/test} &	{\bf More events}  &	{\bf Better constraints~~}  &	{\bf More statistics~~}  	\\%
      \hline
     No-hair ringdown tests    &	$\propto T_{\rm data}$  & $-$	&	$\propto\sqrt{T_{\rm data}}$	\\%
     MBHB  ($n$PN inspiral)         &	$\propto T_{\rm data}$  & $\propto T_\mathrm{elapsed}^{(1+n)/4}$	&	$\propto\sqrt{T_{\rm data}}$	\\%
     SOBHB ($n$PN inspiral)         &	$\propto T_{\rm data}$  & $\propto T_\mathrm{elapsed}^{(1+n)/4}$	&	$\propto\sqrt{T_{\rm data}}$	\\%
     SOBHB (quasicontinuous) &	$\propto T_{\rm data}$ or faster  & $\propto T_{\rm elapsed}^2$	&	$\propto\sqrt{T_{\rm elapsed}}$\\%
     IMR tests                      &	$\propto T_{\rm data}$  & No 	&	$\propto\sqrt{T_{\rm data}}$\\%
     EMRI spacetime mapping   &	$\propto T_{\rm data}$ or faster  & yes	    & $\propto\sqrt{T_{\rm data}}$	\\%
     Ultralight boson bounds    &	$\propto T_{\rm data}^{3/2}$ or 	$\propto T_{\rm data}^{3/4}$	 & marginally 	&	$\propto\sqrt{T_{\rm elapsed}}$	\\%
     Environmental dark matter     &	$\propto T_{\rm data}$ or faster 	 & factor of a few  	&	$-$	\\%
     \hline
     Rare golden events                           &	$\propto T_{\rm data}$  &  factor of a few 	&$-$	 	\\%
     \hline
     \hline
    \end{tabular}
    \caption{Comparison between 4-yr and 6-yr mission duration (applicable to any of the scenarios in Sec.~\ref{sec:intro}) for aspects concerning tests of gravity, of the nature of compact objects, and dark-matter searches with LISA. 
    The first column indicates whether the number of events scales with the actual observing time $T_{\rm data}$; the second column indicates whether we expect better constraints and their scaling with the mission duration time $T_{\rm elapsed}(= T_{\mathrm{data}}/\mathcal{D}$ with a duty cycle $\mathcal{D}$); the third column indicates whether we expect more statistics (e.g., mode stacking, coherent searches, etc.)~\cite{Berti:2011jz,Berti:2016lat,Yang:2017zxs,Perkins:2020tra,Yagi:2009zz}.
    }
    \label{tab:my_label}
\end{table}

A key component of the LISA mission's scientific objectives is to test nature of BHs and search for other dark compact objects~\cite{Cardoso:2019rvt}. 
In particular, elements of SO5 are addressed by investigations of these types, including SI5.1 (``Use ring-down characteristics observed in MBHB coalescences to test whether the post-merger objects are the black holes predicted by GR'') and SI5.2 (``Use EMRIs to explore the multipolar structure of MBHs'').
These investigations share methodologies with tests of the foundations of the gravitational interaction (Sec.~\ref{sec:WP7}) so, as a rule of thumb, we expect the same potential limitations due to a decrease of the effective mission duration.

\subsection{Tests of the nature of black holes}

Here we briefly list the tests of the nature of BHs and searches for compact objects we considered in this study.

\subsubsection{Inspiral-based test with MBHBs, IMBHBs, and EMRIs} 
The sources for these tests are compact binaries in various ranges of masses and mass ratios. The dynamics of these binaries will be affected by {\em dipolar radiation} if the objects are charged (either under an EM or a dark field). It will also be impacted if the {\em multipolar structure} of the binary components differs from that predicted in Kerr, where all multipoles are determined by the mass and spin through elegant relations~\cite{Hansen:1974zz}.
In particular, smoking guns of the non-Kerrness of an object would be the presence of moments that break equatorial symmetry or axisymmetry, as in the case of multipolar boson stars~\cite{Herdeiro:2020kvf} and of fuzzball microstate geometries~\cite{Bena:2020see,Bianchi:2020bxa,Bena:2020uup,Bianchi:2020miz,Bah:2021jno}; or the lack of efficient absorption of radiation by the objects (i.e. {\it tidal heating}), at variance with the BH case. 
For EMRIs in the LISA band, measurements of tidal heating can be used to put a very stringent upper bound on the reflectivity of the object's surface, at the level of $0.01\%$~\cite{Datta:2019epe}. 
In addition, the presence of {\em tidal deformability} effects (other than the aforementioned tidal heating), which are absent for BHs~\cite{Binnington:2009bb,Damour:2009vw} but are generically non-zero for other objects, can leave detectable imprints in the LISA band~\cite{Cardoso:2017cfl,Maselli:2017cmm,Pani:2019cyc}.

\subsubsection{Ringdown tests} 

Measuring the ringdown modes in the post-merger signal of a binary coalescence provides a clean and robust way to the nature of the remnant. 
Detecting several QNMs would allow for multiple independent null-hypothesis (Kerr) tests, and enable GW spectroscopy~\cite{Kokkotas:1999bd,Berti:2009kk}, in particular for golden events~\cite{Dreyer:2003bv,Berti:2016lat}. Besides deforming the QNM spectrum, if the remnant differs from a Kerr BH, some further smoking gun deviations in the prompt ringdown can be the presence of other modes or extra degrees of freedom and the existence of mode doublets arising from isospectrality breaking~\cite{Maggio:2020jml}. 
Even in the absence of deviations in the prompt ringdown, GW echoes~\cite{Cardoso:2016rao,Cardoso:2016oxy,Cardoso:2017cqb} in the late-time post-merger signal of a compact binary coalescence might be a generic smoking gun of new physics at the horizon scale (see~\cite{Cardoso:2019rvt,Abedi:2020ujo} for some recent reviews). The echo amplitude depends on the object's reflectivity~\cite{Mark:2017dnq} that can be constrained only by SNRs of $\mathcal{O}(100)$ in the post-merger phase~\cite{Testa:2018bzd, Maggio:2019zyv}.
This makes LISA particularly well suited for echo searches and gives the tantalizing prospect of probing the near-horizon (possibly quantum) structure of dark compact objects.
Finally, the high sensitivity of LISA could be used  to test proposals for the area quantization of BHs~\cite{Bekenstein:1974jk,Mukhanov:1986me} with suitably modified inspiral--merger--ringdown signals~\cite{Cardoso:2019apo,Brustein:2020tpg,Agullo:2020hxe}.

\subsection{Quantifying the impact of a change in mission duration}

The impact of a change in mission duration depends on the relative magnitude of the signal duration $T_{\rm signal}$ and of the mission duration $T_{\rm elapsed}$ (recall that the actual observing time is $T_{\rm data}={\cal D}\times T_{\rm elapsed}$, where ${\cal D}$ is the duty cycle).
For tests of the nature of BHs, we expect four different scenarios (summarized in Table~\ref{tab:my_label}):

\noindent
{\bf Case~a:} $T_{\rm signal}\ll T_{\rm elapsed}$.
For signals that are short relative to the mission duration, we expect the primary benefit of a longer mission to be the detection of a larger number of signals, with $N_{\rm signal} \propto T_{\rm data}$.
Multiple events can be combined in order to derive constraints on the nature of black holes, and such constraints should obey the usual $1/\sqrt{N_{\rm signal}}$ scaling (in the limit of a large number of similar detections). Thus, for these shorter transients, we expect bounds to improve as $\sqrt{T_{\rm data}}$.
Signals in this category would include MBHBs, which are the primary candidates for no-hair tests with ringdown, for those parametrized inspiral tests which are impacted by properties of the BHs, and for post-merger echo searches of deviations from classical horizons~\cite{Mark:2017dnq,Cardoso:2017cqb,Testa:2018bzd,Abedi:2020ujo}. 

\noindent
{\bf Case~b:} $T_{\rm signal}\gg T_{\rm elapsed}$.
For sources with signals that are long compared to the mission lifetime, increasing the mission duration could have a much stronger impact on the measurements. SOBHBs, when they last a significant portion of the mission duration, fall into this category, as do Galactic binaries which include BHs. The impact of mission duration on constraints for this class of systems depends on the scaling of the phase evolution with time for a given source. For approximately monochromatic sources, a change in the frequency derivative due to non-GR/non-BH effects would result in a phase drift $\propto T_{\rm elapsed}^2$, as discussed in Sec.~\ref{sec:WP2}. For these sources then we expect constraints to scale as $T_{\rm elapsed}^2$, and the number of detections will scale better than $T_{\rm data}$, since quiet signals can accumulate SNR over the entire observational data $T_{\rm data}$. 

{\noindent
\bf Case~c:} $T_{\rm signal}\sim T_{\rm elapsed}$.
EMRI events are the most representative example of an intermediate case, and are particularly relevant for tests of the nature of supermassive objects since they can potentially provide unparalleled constraints. EMRIs can last a significant amount of time, and so we expect that the number of EMRI detections will improve faster than linearly with $T_{\rm data}$, as discussed in Sec.~\ref{sec:WP3}.
A simple way to estimate the impact on mission duration on the detection of these sources is to require that the system be observed for at least some amount of observation time $T_0$ before it can be used.
Then the amount of time during which these signals can actually be detected is $T_{\rm det} = T_{\rm data} - T_0$. By increasing $T_{\rm data}$ by a factor $\gamma$, we see that $T_{\rm det} \to \gamma T_{\rm det} +  T_0(\gamma -1)$.
This results in an increase in the number of detections which is linear in $\gamma$, but with an additive factor. The lowest mass MBHBs will also take a significant amount of time to inspiral, and are covered by this intermediate case.

\noindent
{\bf Case~d:} Rare golden events.
Finally, for certain scientific goals and especially for precision tests of gravity and of the nature of BHs, rare golden events can make a major difference, since they are paramount for major and groundbreaking discoveries. The probability of detecting one or more rare events scales approximately with the amount of time observed, so that the expected number of such rare events also scales linearly with $T_{\rm data}$.

We conclude that SO5 of the LISA mission proposal, ``Explore the fundamental nature of gravity and black holes,'' would be facilitated by a longer mission duration, with the expected number of events (including rare golden events of paramount importance for fundamental physics) increasing linearly with mission duration. In some cases, especially for long-duration signals, we expect better than linear improvement in the number of detected events and/or in the constraints derived from each event.

\section{Conclusions}
\label{sec:summary}

In this paper we have examined the performance of the various scenarios described in the introduction with respect to the LISA SOs defined in the mission proposal~\cite{2017arXiv170200786A} for the configuration SciRD. An in-depth scrutiny of the scientific capabilities of LISA has revealed that the adopted mission duration has a strong impact on several SOs and SIs, as defined in the LISA proposal. Although all areas of LISA science (astrophysics, cosmology, and fundamental physics) are affected to some extent, the impact is more prominent for some of the astrophysics goals.

\setlength{\tabcolsep}{-0.1pt} 
\begin{table}[t]
\centering
\vspace*{-\baselineskip}
\begin{tabular}{| p{5.5cm} | p{1.0cm} p{1.0cm} p{1.0cm} | p{1.0cm} | p{1.0cm} p{1.0cm} p{1.0cm} | }
\hline
Scenario & T4C & T4G5 & T4G1 & T5C & T6C & T6G5 & T6G1 \\
\hline
$T_{\rm elapsed}$&
\multicolumn{3}{c|}{$4$\,yr}&
$5$\,yr&
\multicolumn{3}{c|}{$6$\,yr}\\
$T_{\rm data}=0.75\times T_{\rm elapsed}$&
\multicolumn{3}{c|}{$3$\,yr}&
$3.75$\,yr&
\multicolumn{3}{c|}{$4.5$\,yr}\\
\hline
Gaps & one & 5 days & 1 day & one & one & 5 days & 1 day \\
\hline
Galactic binaries (\small{SO1~SI1.2}) (\S\ref{sec:WP2})  &\ycel  &\ycel  &\ycel &\gcel &\bcel &\bcel &\bcel\\
Black hole seeds (\small{SO2~SI2.1}) (\S\ref{sec:WP1})   &\ycel  &\rcel  &\rcel &\ycel &\gcel &\gcel &\gcel\\
EM counterparts (\small{SO2~SI2.3}) (\S\ref{sec:WP1}, \S\ref{sec:WP4})  &\ycel  &\ycel  &\ycel &\gcel &\gcel &\gcel &\gcel\\
EMRIs (\small{SO3~SI3.1})  (\S\ref{sec:WP3}) &\ycel  &\ycel  &\ycel &\gcel &\bcel &\bcel &\bcel\\
Multiband SOBHs (\small{SO4~SI4.1}) (\S\ref{sec:WP2})  &\rcel  &\rcel  &\rcel &\ycel &\gcel &\gcel &\gcel\\
SOBH formation (\small{SO4~SI4.2}) (\S\ref{sec:WP2})  &\rcel  &\rcel  &\rcel &\ycel &\gcel &\gcel &\gcel\\
Kerr tests (\small{SO5~SI5.1\&5.2}) (\S\ref{sec:WP8})  &\ycel  &\ycel  &\ycel &\ycel &\gcel &\gcel &\gcel\\
Tests of GR (\small{SO5~SI5.3\&5.4}) (\S\ref{sec:WP7})  &\ycel  &\ycel  &\ycel &\ycel &\gcel &\gcel &\gcel\\
Ultralight bosons (\small{SO5~SI5.5}) (\S\ref{sec:WP6})  &\ycel  &\ycel  &\ycel &\ycel &\gcel &\gcel &\gcel\\
$H_0$ via standard sirens (\small{SO6~SI6.1}) (\S\ref{sec:WP5})  &\ycel  &\wcel  &\wcel &\gcel &\gcel &\wcel &\wcel\\
Cosmological parameters (\small{SO6~SI6.2}) (\S\ref{sec:WP5})  &\ycel  &\ycel  &\ycel &\ycel &\ycel &\gcel &\gcel\\
\hline
\end{tabular}
\caption{\label{tab:I} List of SOs and SIs that are degraded when a duty cycle ${\cal D}=0.75$ is applied to the baseline LISA mission, defined as SciRD. SOs are listed in ascending order following the LISA proposal~\cite{2017arXiv170200786A}. The hyperlinks in parentheses next to each SO refer to the sections of the present document used to draw the conclusions summarized in this table. The different colors (red, green, yellow, blue) indicate whether each SO/SI goal is met, according to the interpretation provided in the main text. In the definition of gaps, ``one'' means that the data set is only reduced by a factor of ${\cal D}=0.75$ relative to $T_{\rm elapsed}$ as a consequence of a single long gap {\em either at the beginning or at the end of the mission}: for example, in scenario T4C we have $T_{\rm data}={\cal D}\times T_{\rm elapsed}=3$~yr of continuous data. White entries appear because for SI6.1 we could not study the effect of gaps.}
\end{table}

Our main findings are summarized in Table~\ref{tab:I}, where the color code has the following interpretation:
\begin{itemize}
    \item {\bf green}: the objective, as defined in the LISA proposal for SciRD, can be achieved;
    \item {\bf yellow}: we cannot establish whether the objective can be achieved, because of astrophysical uncertainties, or because the results would need deeper verification. Nonetheless, our investigation points towards a substantial performance degradation compared to SciRD;
    \item {\bf red}: there is a significant danger of failing the objective as defined for SciRD;
    \item {\bf blue}: there is an improvement in the capabilities of the instrument compared to SciRD.
\end{itemize}

In the table we only list the SOs for which configuration T4C (i.e.\ a reduction in the usable data stream due to the ${\cal D}=0.75$ duty cycle) corresponds to either a degradation of the SO (yellow) or danger of failing the goals stated in the LISA proposal (red).

Based on the analysis presented in this paper, we strongly recommend an extension to 6~yr of mission operation. The recommendation is based on the following assessment of the impact of mission duration on individual LISA SOs. 

\begin{itemize}
    \item {\bf {SO1. SI1.1:} Enable joint gravitational and EM observations of Galactic binaries to study the interplay between gravitational radiation and tidal dissipation in interacting stellar systems.}\\
The study of this interplay relies on the measurement of the frequency derivatives of the GW signal, to discriminate GW vs mass transfer driven evolution, unveil tidal interactions, etc. 
The number of Galactic binaries for which $\dot{f}$ and $\ddot{f}$ can be measured scales with $T_{\rm elapsed}^2$ and $T_{\rm elapsed}^3$, respectively. The benefits of extending the mission to 6~yr are therefore clear, especially when considering that measuring $\ddot{f}$ will be feasible only for a handful of sources.

\item {\bf {SO2. SI2.1:} Search for seed black holes at cosmic dawn.}\\ 
Inclusion of gaps in the data stream for a total duty cycle ${\cal D}=0.75$ significantly affects the number of observable high-redshift ($z>10$), low-mass ($M<10^3\msun$) MBHBs. In our standard models, assuming a 4-yr mission, those are reduced from $\approx25$ for SciRD to $\lesssim 10$ for the scenarios with gaps in the data (T4G5 and T4G1). For more pessimistic scenarios, the number of low-mass MBHB detections decreases from $\approx 10$ to $\lesssim 6$. 
Those are dangerously low numbers that can jeopardize our ability to reconstruct the nature of the first MBH seeds. Extending the mission to 6~yr would put this investigation on safer ground, increasing the low-mass/high-redshift seed MBH sample from 10 to 15 in our standard model. This is necessary in order to address SI2.1.

\item {\bf {SO2. SI2.3:} Observation of EM counterparts to unveil the astrophysical environment around merging MBHBs.}\\
  The number of sources at $z<2$ which are primary targets for EM follow-ups are expected to be just a few ($\approx 2$ yr$^{-1}$ in our fiducial models). Compared to SciRD, the presence of a 0.75 duty cycle will severely degrade the SNR and sky localization of $\approx 30\%$ of these sources, posing a significant threat to the success of associated EM searches. %
It is therefore essential to extend the mission to 6~yr, which mitigates the risk of failing SI2.3, since the number of detected massive and nearby sources scales with $T_{\rm elapsed}$. 

\item {\bf {SO3. SI3.1:} Study the immediate environment of Milky Way-like MBHs at low redshift.}\\
The presence of gaps in the data will make it harder to observe EMRIs up to $z\approx 4$, which is the goal stated in the LISA proposal. Our simulations indicate that the number of observable EMRIs scales with $\approx T_{\rm data}^{3/2}$, roughly doubling the number of observed systems for a mission duration extension from 4~yr to 6~yr. This will mitigate the chances of missing EMRIs altogether should we face the most pessimistic astrophysical scenarios, which forecast $\approx 1$ observable EMRI per year. The SNR for detecting deviations in EMRI waveforms due to environmental effects (e.g., the SOBH's interaction with circumbinary gas) scales more steeply with mission duration, as $\propto T^{2}$--$T^3$~\cite{Kocsis:2011dr}, further justifying the extension to a 6~yr mission. 

\item {\bf {SO4. SI4.1:} Study the close environment of stellar-origin black holes (SOBHs) by enabling multi-band and multi-messenger observations at the time of coalescence.}\\
The inclusion of gaps in the data, together with the relaxed high-frequency sensitivity requirement (by a factor 1.5 compared to the LISA proposal design) pose major obstacles to the fulfillment of this objective. With 4~yr of observations and ${\cal D}=0.75$ duty cycle (i.e. $T_{\rm data}=3$ yr, as per T4C, T4G5, and T4G1), the expectation is to observe a couple of multi-band sources, and the SNR $>8$ goal 
on GW150914-like sources is difficult to achieve. A 6-yr extension will double the number of multi-band systems, which is crucial for SI4.1.

\item {\bf {SO4. SI4.2:} Disentangle SOBH binary formation channels.}\\
For the reasons mentioned above, the number of detectable SOBHs is likely going to be ${\cal O}(10)$, which might be insufficient to statistically discriminate formation channels via eccentricity measurements. Since the number of observable SOBHs also scales with $\approx T_{\rm data}^{3/2}$, an extension to 6~yr will double the number of detections, allowing for a better measurement of the eccentricity distribution, which is of paramount importance for SI4.2.

 \item {\bf {SO5. SI5.2:} Use of EMRIs to test multipolar structure.}\\
Mapping the spacetime around a BH using an EMRI signal is not endangered by a mission duration of 4~yr {\em if EMRIs are observed} (cf.\ SI3.1 above), but weak EMRI signals will build up throughout the mission duration. A longer mission thus results in improvements that scale faster than linear with the mission lifetime for these tests. 

 \item {\bf {SO5. SI5.3 and SI5.4:} Propagation properties of GWs and other emission channels.}\\
Many fundamental questions in gravitational physics, such as the dispersion effects induced by a nonzero graviton mass, the existence of dipolar charges, a time-varying Newton's constant, and environmental effects due (say) to dark matter, can be addressed jointly via a parametrized formalism. Mission duration has an impact on our ability to constrain these parameters, especially when they affect the waveform at low frequencies, and require long observation times to remove degeneracies. A rough scaling of these bounds on the associated ppE coefficients with mission duration is given in Eq.~\eqref{eq:beta_T_scaling}: for example, the bounds on environmental and dark matter effects will degrade by up to a factor of two if the mission lifetime is reduced from 6~yr to 4~yr. 

For the reasons highlighted above (SO2. SI2.3), we may also miss several golden events, and this would affect BH spectroscopy tests based on the detection of multiple harmonics of the ringdown.

\item {\bf {SO5. SI5.5:} Test the existence of ultralight fields and discover dark matter spikes.}\\
Ultralight fields can produce monochromatic GW signals through superradiance. The mission duration has a significant impact on the number of resolvable sources of such monochromatic GWs, which scales super-linearly (cf.\ Table~\ref{tab:my_label}). Therefore, mission duration affects our ability to discover ultralight dark matter. It also impacts the constraints on the local dark matter density in some binaries, with up to a factor of 2 improvement if the mission is extended from 4~yr to 6~yr.
 
 \item {\bf {SO6. SI6.1 and SI6.2:} Probe the rate of expansion of the Universe.}\\
Different categories of sources enable LISA to probe the expansion of the Universe at different redshift. In particular, SI6.1 selects SOBHB and EMRIs as distance indicators, to probe the Hubble parameter today with statistical identification of the redshift. Preliminary results using EMRIs alone seem to indicate that a 5-yr mission with ${\cal D}=0.75$ (i.e.\ configuration T5C) is the minimum necessary to constrain the Hubble parameter today to better than 2\% (SI6.1). SI6.2 selects MBHB as distance indicators, with redshift identification coming from an EM counterpart. We propose a new FoM to meet this science objective, i.e.\ the measurement of the Hubble rate at redshift 2. In the most pessimistic astrophysical scenario for the MBHB formation channel, the FoM cannot be met, but it can be met for two more optimistic scenarios. 

\item {\bf {SO7. SI7.1\&7.2:} Understand stochastic GW backgrounds and their implications for the early Universe and TeV-scale particle physics.}\\
While extending the overall mission duration would improve the science return of LISA concerning SO7, both SI7.1 and SI7.2 can be met with 3~yr of continuous data. Gaps are not expected to affect the detection of a stochastic GW background.

\end{itemize}

In summary, the introduction of a 75\% duty cycle on a 4-yr mission duration (i.e.\ configurations T4C, T4G5, and T4G1) has a detrimental effect on several of the SOs and SIs that are the foundation of the LISA science case. 

A mission extension to 6~yr (configurations T6C, T6G5, and T6G1) will:
\begin{itemize}
    \item mitigate the risk of failing SI2.1, SI2.3, SI4.1, SI4.2, SI6.2;
    \item be beneficial for SI1.1, SI3.1, SI6.1 and all investigations related to SO5, with significant improvement for SI5.2, SI5.5, and some parametrized constraints. Especially for SI1.1 and SI3.1, even though they are not at risk under assumptions T4C, T4G5, and T4G1, we stress that the number of sources that can be used to address these SIs increases more than linearly with the duration of the mission.
\end{itemize}

\begin{acknowledgements}

We thank Danny Laghi and Walter Del Pozzo for help with Fig.~\ref{fig:EMRIsH0}.
C.P.L.~Berry is supported by the CIERA Board of Visitors Professorship. 
E.~Berti is supported by NSF Grants No. PHY-1912550 and AST-2006538, NASA ATP Grants No.\ 17-ATP17-0225 and 19-ATP19-0051, NSF-XSEDE Grant No. PHY-090003, and NSF Grant PHY-20043. 
D.~Blas acknowledges support from the Fundaci\'on Jesus Serra and the Instituto de Astrof\'isica de Canarias under the Visiting Researcher Programme 2021 agreed between both institutions.
T.~Bogdanovi\'c acknowledges the support by the National Aeronautics and Space Administration (NASA) under award No. 80NSSC19K0319 and by the National Science Foundation (NSF) under award No.\ 1908042.
V.~Cardoso acknowledges financial support provided under the European Union's H2020 ERC Consolidator Grant ``Matter and strong-field gravity: New frontiers in Einstein's  theory'' grant agreement no.\ MaGRaTh--646597.
This project has received funding from the European Union's Horizon 2020 research and innovation programme under the Marie Sklodowska-Curie grant agreement No 101007855.
We thank FCT for financial support through Projects~No.~UIDB/00099/2020 and through grants PTDC/MAT-APL/30043/2017 and PTDC/FIS-AST/7002/2020.
H.-Y.~Chen is supported by NASA through NASA Hubble Fellowship grants No.\ HST-HF2-51452.001-A awarded by the Space Telescope Science Institute, which is operated by the Association of Universities for Research in Astronomy, Inc., for NASA, under contract NAS5-26555. 
Z.~Haiman acknowledges support by NASA grant NNX15AB19G and NSF grants AST-2006176 and AST-1715661.
G.~Nardini~is partly supported by the ROMFORSK  grant  project.~no.~302640  “Gravitational  Wave  Signals  From  Early Universe Phase Transitions”.
P.~Pani acknowledges financial support provided under the European Union's H2020 ERC, Starting Grant agreement no.~DarkGRA--757480. He also acknowledges support under the MIUR PRIN and FARE programmes (GW-NEXT, CUP: B84I20000100001), and by the Amaldi Research Center funded by the MIUR program ``Dipartimento di Eccellenza'' (CUP: B81I18001170001).
A.~Sesana acknowledges financial support provided under the European Union’s H2020 ERC Consolidator Grant ``Binary Massive Black Hole Astrophysics'' (B Massive, Grant Agreement: 818691). K.~Yagi acknowledges support from NSF Grant PHY-1806776, NASA Grant 80NSSC20K0523, a Sloan Foundation Research Fellowship and the Owens Family Foundation.
D.~J.~Weir was supported by a Science and Technology Facilities Council Ernest Rutherford Fellowship, Grant no. ST/R003904/1, and by the Academy of Finland, Grants 324882 and 328958.
A.~Zimmerman is supported by NSF Grant No.~PHY-1912578.
The authors would like to acknowledge networking support by the GWverse COST Action CA16104, ``Black holes, gravitational waves and fundamental physics.''
The Flatiron Institute is supported by the Simons Foundation.
This research has made use of data, software and web tools obtained from the Gravitational Wave Open Science Center (\href{https://www.gw-openscience.org/}{www.gw-openscience.org/}), a service of LIGO Laboratory, the LIGO Scientific Collaboration and the Virgo Collaboration. LIGO Laboratory and Advanced LIGO are funded by the United States National Science Foundation (NSF) as well as the Science and Technology Facilities Council (STFC) of the United Kingdom, the Max-Planck-Society (MPS), and the State of Niedersachsen/Germany for support of the construction of Advanced LIGO and construction and operation of the GEO600 detector. Additional support for Advanced LIGO was provided by the Australian Research Council. Virgo is funded, through the European Gravitational Observatory (EGO), by the French Centre National de Recherche Scientifique (CNRS), the Italian Istituto Nazionale di Fisica Nucleare (INFN) and the Dutch Nikhef, with contributions by institutions from Belgium, Germany, Greece, Hungary, Ireland, Japan, Monaco, Poland, Portugal, Spain.
\end{acknowledgements}

\begin{table}[ht]
\caption{List of acronyms.}
\label{tab:acronyms}
 \begin{tabular}{|p{2.0cm}|p{9.5cm}|}
   \hline
    BBH & binary black hole\\
    ${\cal D}$ & duty cycle\\
    DM & dark matter\\
    DWD & double white dwarf\\
    EM & electromagnetic\\
    EMRI & extreme mass ratio inspiral\\
    FOPT & first order phase transition\\
    GR & general relativity\\
    GW & gravitational wave\\
    IMRI & intermediate mass ratio inspiral\\
    $\Lambda$CDM & standard cosmology with cold dark matter and cosmological constant\\ 
    LISA & Laser Interferometer Space Antenna\\
    MBH & massive black hole\\
    MBHB & massive black hole binary\\
    PBH & primordial black hole\\
    PN & post-Newtonian\\
    popIII & population III stars/light-seed BH model\\
    ppE & parametrized post-Einsteinian\\
    Q3d & heavy-seed black hole model with delays\\
    Q3nd & heavy-seed black hole model with no delays\\
    QNM & quasi-normal mode\\
    SciRD & Science Requirement Document\\
    SGWB & stochastic gravitational wave background\\
    SI & science investigation\\
    SNR & signal-to-noise ratio\\
    SO & science objective\\
    SOBH & stellar origin black hole\\
    SOBHB & stellar origin black hole binary\\
    $T_{\rm data}$ & length of data for scientific use\\ 
    $T_{\rm elapsed}$ & nominal mission duration\\
    $T_{\rm signal}$ & signal lifetime in the LISA band\\
    T4C & 3 yr of continuous scientific data\\
    T5C & 3.75 yr of continuous scientific data\\
    T6C & 4.5 yr of continuous scientific data\\
    T4G1 & 3 yr of scientific data with 1 day gaps\\
    T4G5 & 3 yr of scientific data with 5 day gaps\\
    T6G1 & 4.5 yr of scientific data with 1 day gaps\\
    T6G5 & 4.5 yr of scientific data with 5 day gaps\\
    TDI & time-delay interferometry\\
    XMRI & extremely large mass ratio inspiral\\
    \hline
\end{tabular}
\end{table}

\bibliographystyle{spphys}
\bibliography{References}

\end{document}